\documentclass[%
 reprint,
 superscriptaddress,
 amsmath,amssymb,
 pre,
floatfix,
]{revtex4-2}

\usepackage{color}

\usepackage{float}
\usepackage{graphicx}% Include figure files
\usepackage{dcolumn}% Align table columns on decimal point
\usepackage{bm}% bold math
\usepackage{hyperref}% add hypertext capabilities
\usepackage[dvipsnames]{xcolor} %Allows different textcolors
\usepackage{ulem}  %Allows strikethrough
\DeclareMathOperator\erf{erf}

\newcommand{\RN}[1]{%
  \textup{\uppercase\expandafter{\romannumeral#1}}%
}
\begin{document}

%\preprint{APS/123-QED}

\title{Competition between lanes and transient jammed clusters in driven binary mixtures}

\author{Honghao Yu}
\affiliation{%
 Yusuf Hamied Department of Chemistry, University of Cambridge, Lensfield Road, Cambridge CB2 1EW, United Kingdom}%
\author{Robert L. Jack}%
\affiliation{%
 Yusuf Hamied Department of Chemistry, University of Cambridge, Lensfield Road, Cambridge CB2 1EW, United Kingdom}%
\affiliation{%
 Department of Applied Mathematics and Theoretical Physics, University of Cambridge, Wilberforce Road,
Cambridge CB3 0WA, United Kingdom}%

\newcommand{\beq}{\begin{equation}}
\newcommand{\eeq}{\end{equation}}
\newcommand{\CorrectText}[2]{{\sout{#1}}{~{#2}}}
\normalem % ulem makes emph into an underline, so this fixes that

\newcommand{\Pe}{\mathrm{Pe}}

\date{\today}% It is always \today, today,

\begin{abstract}
We consider mixtures of oppositely driven particles, showing that their non-equilibrium steady states form lanes parallel to the drive, which coexist with transient jammed clusters where particles are temporarily immobilised.  We analyse the interplay between these two types of non-equilibrium pattern formation, including their implications for macroscopic demixing perpendicular to the drive.  Finite-size scaling analysis indicates that there is no critical driving force associated with demixing, which appears as a crossover in finite systems.  We attribute this effect to the disruption of long-ranged order by the transient jammed clusters.
\end{abstract}

\maketitle

\section{Introduction}

Non-equilibrium systems can form complex patterns in their steady states, including flocking of birds~\cite{couzin2003self, ballerini2008empirical, bialek2012statistical}, active nematics~\cite{sanchez2012spontaneous, doostmohammadi2018active}, and motility-induced phase separation~\cite{tailleur2008statistical, thompson2011lattice, cates2015motility}.  These patterns are sustained by the continuous injection of energy into the system, which is dissipated as heat.  The formation and evolution of large coherent structures can sometimes be captured by deterministic equations~\cite{turing1952chemical, gray1983autocatalytic, burger2016lane}, but some chaotic fluctuating patterns necessitate a stochastic description, using methods of non-equilibrium statistical physics.  
These systems challenge existing theories: How should emergent patterns be measured and classified? Can one identify distinct dynamical phases, analogous to equilibrium systems?  Can the patterns be controlled by external perturbations?  

A common example of a fluctuating pattern is the emergence of \emph{lanes}, when two species of particles are driven in opposite directions: particles of the same species tend to follow each other.   In physics, this occurs for colloidal systems in electric fields~\cite{dzubiella2002lane, leunissen2005ionic, vissers2011lane}, and in plasmas~\cite{sutterlin2009dynamics, sutterlin2009lane, du2012experimental, sarma2020lane}, and ionic liquids~\cite{kondrat2014charging}.  Similar structures are also familiar from pedestrian flow~\cite{helbing1995social, isobe2004experiment, karamouzas2014universal, oliveira2016keep, bacik2023lane}, and ant foragers~\cite{couzin2003self1}, and oscillatory driving can also induce other patterns~\cite{wysocki2009oscillatory, vissers2011band, li2021phase}.  

Laning behaviour can be captured in computer simulations of model systems where particles interact by repulsive short-ranged potentials and move by Brownian dynamics.  
Experiments on laning involve more complex interactions, such as electrostatic or hydrodynamic interactions in colloids~\cite{leunissen2005ionic, vissers2011lane, vissers2011band}, or non-trivial decision-making by pedestrians~\cite{helbing1995social, isobe2004experiment, karamouzas2014universal, oliveira2016keep}.  Nevertheless, simple models can capture generic laning behaviour, indicating that this 
robust collective phenomenon~\cite{dzubiella2002lane, rex2008influence, glanz2012nature, kohl2012microscopic, klymko2016microscopic, poncet2017universal} is amenable to the methods of statistical physics.

A generic mechanism for laning is enhanced lateral diffusion (ELD)~\cite{chakrabarti2003dynamical, chakrabarti2004reentrance, reichhardt2006cooperative, klymko2016microscopic, vasilyev2017cooperative, schimansky2021demixing, yu2022lane, vansaders2023informational}: collisions between oppositely driven particles promote rapid diffusive motion perpendicular to the field.  This favours accumulation in lanes, where such collisions are avoided.  Despite considerable work on these non-equilibrium steady states, open questions remain about their generic properties, such as the possible existence of long-ranged order, and de-mixing of the system into domains, perpendicular to the drive~\cite{glanz2012nature, kohl2012microscopic, klymko2016microscopic, geigenfeind2020superadiabatic, yu2022lane}.

This paper presents computer simulations of a model system that exhibits laning.  
The main phenomenology is outlined in Fig.~\ref{fig:Configuration}, with details given below.
Red and blue particles are driven in opposite directions by a force that is parameterised by the Peclet number (Pe).
 Increasing Pe from zero, one observes increasingly inhomogeneous steady states (labelled \RN{2}-\RN{4}), which may be contrasted with the homogeneous equilibrium mixture at $\Pe=0$ (state \RN{1}).  
 We find three main results:
(i) Using extensive simulations for different system sizes and driving strength, we characterise for the interplay between the laning state \RN{3} and the demixed state \RN{4}. (ii) While state \RN{4} occurs in finite systems, we show that sufficiently large systems are always mixed: hence demixing is not a phase transition.
(iii) We characterise the large density fluctuations in states \RN{2} and \RN{3} in terms of transient jammed clusters (TJCs), which tend to disrupt the lanes, suppressing long-ranged order.
Previous work~\cite{glanz2012nature} analysed states $\RN{2}$ and $\RN{3}$ by focussing on long-ranged order parallel to the drive; here we analyse phase separation perpendicular to the drive, showing that state $\RN{4}$ is suppressed in large systems.
Hence, we interpret  
laning behaviour in these systems as a combination of ELD (favouring lane formation), and transient clustering (disrupting laning).

\begin{figure}[t]
    \centering 
    \includegraphics[width=0.48\textwidth]{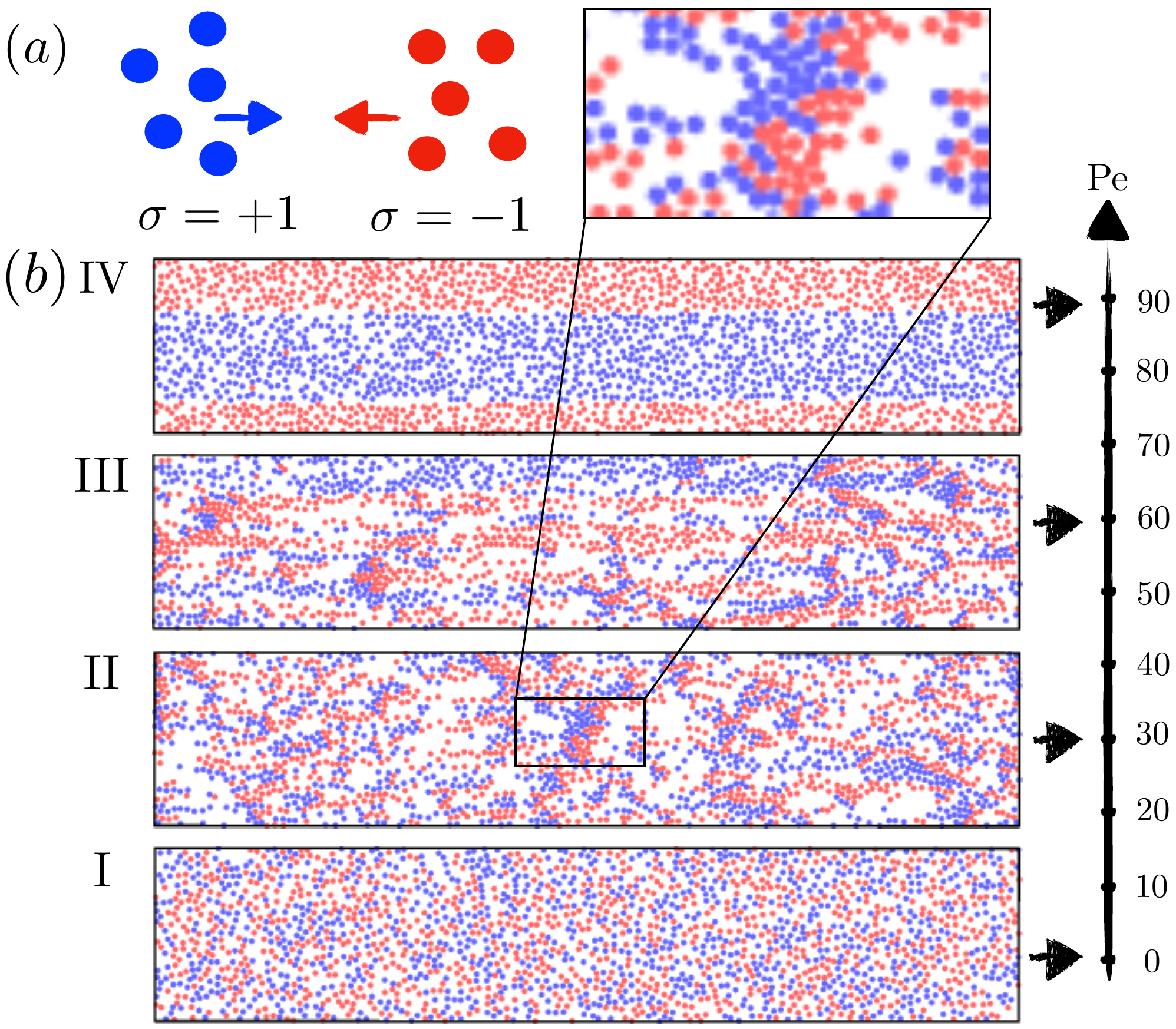}
    \caption{(a)~Example of a transient jammed cluster. (b)~Steady state configurations of a system with $L_{\parallel}=150d_0$, varying $\Pe$, as shown.  Supplemental Material movies 1-3 show trajectories from  steady states \RN{2}-\RN{4}~\cite{SM}. 
    }
    \label{fig:Configuration}
\end{figure}

\section{Model}

{To capture the generic phenomenology of laning -- as it appears in a variety of practical contexts~\cite{dzubiella2002lane, rex2008influence, glanz2012nature, kohl2012microscopic, klymko2016microscopic, poncet2017universal} -- we adopt a simplified model that is sufficient to capture the behavior of interest.  Similar to previous works~\cite{dzubiella2002lane, kohl2012microscopic, glanz2012nature, klymko2016microscopic, oliveira2016keep, poncet2017universal, geigenfeind2020superadiabatic, bacik2023lane}, the model captures the fact that two species of particle are driven in opposite directions and that they feel strong repulsive interactions that prevent them from overlapping.  Their motion is biased by the driving field but also includes a noise term, which can describe Brownian motion in colloidal systems or other random influences from the environment.  Specific laning systems might include other interactions, such as electrostatic forces in colloids or more complex decision rules for pedestrians, but these are not generically required for laning to occur, so we neglect them, consistent with~\cite{dzubiella2002lane, kohl2012microscopic, glanz2012nature, klymko2016microscopic, oliveira2016keep, poncet2017universal, geigenfeind2020superadiabatic, bacik2023lane}.}

{To this end,} we consider a binary mixture of $N$ particles, with $N/2$ in each species, occupying a two dimensional box of size $L_{\parallel}\times L_{\perp}$, with periodic boundary conditions.  {The species of particle $i$ is labelled as $\sigma_i=+1$ (blue) or $\sigma_i =-1$ (red).}  
Particles are driven by a 
{force} of strength $E$ along the $x$-axis and interact with each other via WCA potentials~\cite{weeks1971role}, 
\begin{equation}
V(r)= 4\epsilon[(d_0/r)^{12} - (d_0/r)^{6} + (1/4) ]\Theta(2^{1/6}d_0-r),
\end{equation}
where $d_0$ is the diameter, $\epsilon$ is the repulsion strength, and $\Theta$ is the Heaviside function. 
The position $\bm{r}_i=(x_i,y_i)$
of particle $i$ follows Langevin dynamics, implemented in LAMMPS~\cite{LAMMPS},
\beq\label{eq:Langevin}
m\ddot{\bm{r}}_{i}= - \gamma \dot{\bm{r}}_i - \nabla_i U
+  \sigma_i E \bm{\hat{x}}  + \sqrt{2\gamma k_B T }\, \boldsymbol{\eta}_i,
\eeq
where $\gamma$ is a friction constant, $\bm{\hat{x}}$ is a unit vector in the $x$-direction, $T$ is the temperature of the heat bath, $k_B$ is Boltzmann's constant, $U$ is the interaction energy, and $\bm{\eta}_i$ is a Gaussian white noise with mean zero and 
$\langle {\eta}_{i,\alpha}(t) {\eta}_{j,\beta}(t') \rangle = \delta_{ij}\delta_{\alpha\beta}\delta(t-t')$, where $\alpha,\beta$ indicate Cartesian components.
The Brownian time is $\tau_{\rm B}=d_0^2\gamma/(k_B T)$.

\newcommand{\asp}{S}

We non-dimensionalise the system
using base units $d_0,m,\epsilon$, see Appendix~\ref{app:model} for full details.  The dimensionless control parameters of the model are the Peclet number $\text{Pe}= E d_0/(k_B T)$, the area fraction $\phi=N\pi d_0^2/(4L_\parallel L_\perp)$, the reduced friction $\tilde\gamma=\gamma d_0/\sqrt{m\epsilon}$, the reduced temperature $\tilde T=k_{\rm B}T/\epsilon$ and the aspect ratio of the simulation box $\asp=L_\parallel/L_\perp$.
We take $\tilde\gamma=1000$ to mimic a high-friction colloidal environment, and $\tilde T=1$ (results depend weakly on this parameter).  
Since particles explore space much more quickly in the $x$-direction, we take $\asp=5$, except where explicitly stated otherwise.  We focus on $\phi=0.35$ which is representative of a laning regime {$0.1\lesssim\phi\lesssim0.4$, some results for other densities are presented in the Appendices.

\begin{figure*}[t]
    \centering 
    \includegraphics[width=0.9\textwidth]{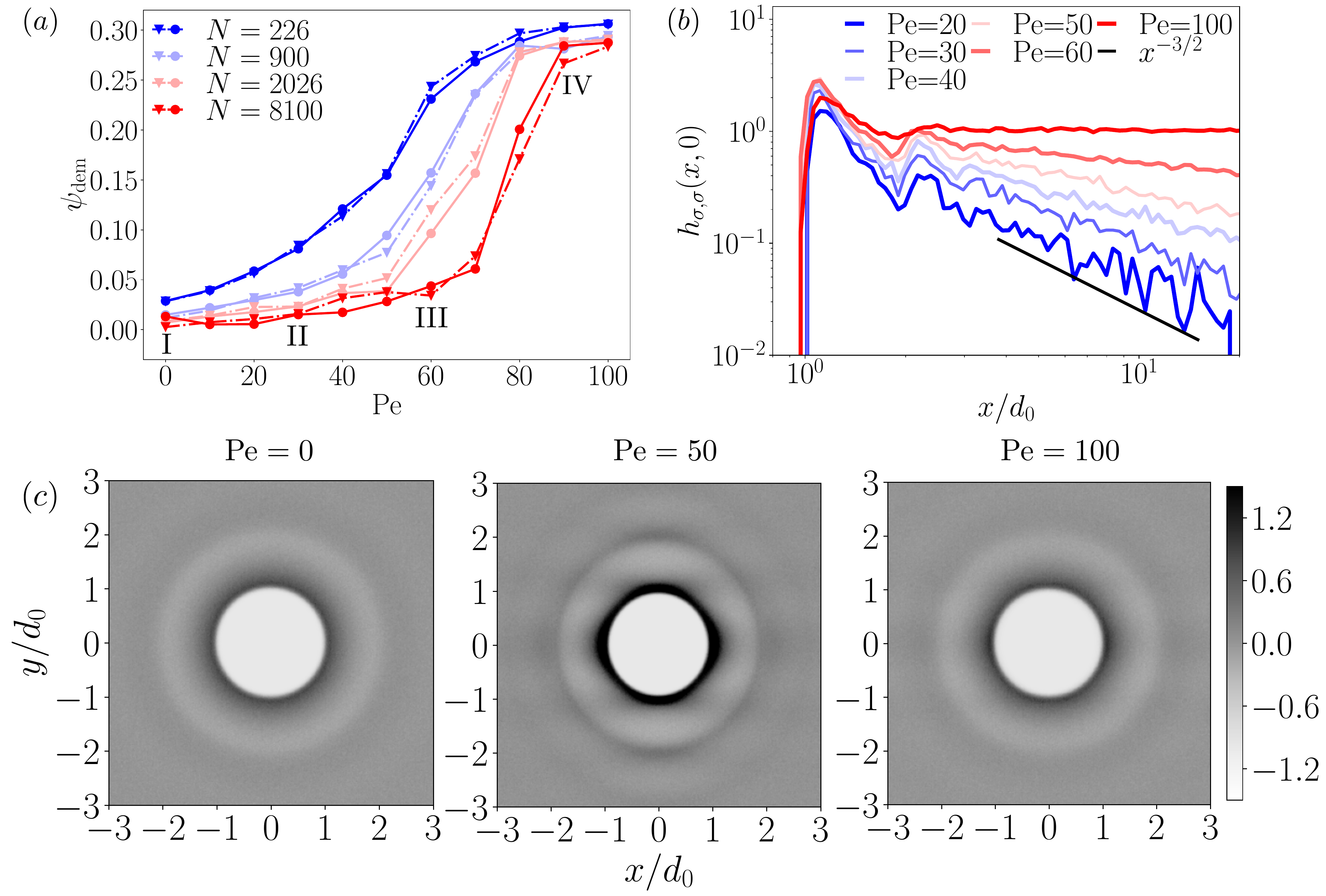}
    \caption{(a)~$\psi_{\rm dem} $ against $\text{Pe}$ at different system sizes.  Dotted and solid lines are obtained from disordered and demixed initial conditions, respectively.
    (b)~$h_{\sigma,\sigma}(\bm{r})$ plotted along the $x$-direction.  The solid line indicates $h_{\sigma,\sigma}( x,0)\sim x^{-3/2}$. 
    (c)~Dependence of $h_{\rho,\rho}(\bm{r})$ on $\bm{r}=(x,y)$.   In (b), (c), $L_\parallel=150d_0$.
    }
    \label{fig:OP}
\end{figure*}

\section{Results}

\subsection{Steady states:  Demixing is a smooth crossover}

The steady state behaviour of this model is illustrated in Fig.~\ref{fig:Configuration}(b), for a range of Pe.
We verified that all systems have converged to their steady states by starting independent simulations from both homogeneous and  fully de-mixed initial conditions, which lead to the same results.  This requires long simulations, up to $4000\tau_B$ (see Appendix~\ref{app:model} for further details of initialisation and equilibration).

\newcommand{\rhobar}{\overline{\rho}}
\newcommand{\psidem}{\psi_{\rm dem}}

To characterise the fully demixed state, {we interpret $\sigma_i$ as a dimensionless charge} and evaluate the Fourier transform of the charge density at $\bm{k}^*=(0,2\pi/L_\perp)$~\cite{korniss1995novel, korniss1997nonequilibrium, yu2022lane}:
\begin{equation}\label{eq:OrderParameter}
\psi_{\rm dem} 
= \langle|\Psi({\bm{k}}^*)|\rangle 
=\Big\langle\Big|\frac{1}{L_{\perp} L_{\parallel}}\sum_{i=1}^{N} \sigma_{i} e^{-i\bm{k}^*\cdot \bm{r}_i}\Big|\Big\rangle,
\end{equation}
Here and throughout, angle brackets indicate averages in the steady state of the dynamics.  

We use $\psi_{\rm dem}$ as an order parameter for demixing, in a finite-size scaling analysis, by increasing $N, L_\parallel, L_\perp$, at fixed $\Pe,\phi$, and aspect ratio $\asp$.  A demixing phase transition would manifest as a critical Peclet number $\Pe^*$:
For $\Pe<\Pe^*$ the state would be homogeneous and $\Psi({\bm{k}}^*)$ would be small in modulus (of order $N^{-1/2}$ as $N\to\infty$).
For $\Pe>\Pe^*$ the state would be demixed: the complex number $\Psi({\bm{k}}^*)$ has a modulus of order $1$; its phase is random and indicates the relative positions of the two domains.  Hence, demixing involves spontaneous symmetry breaking, via the phase of $\Psi$.

\begin{figure*}[t]
    \centering 
    \includegraphics[width=0.95\textwidth]{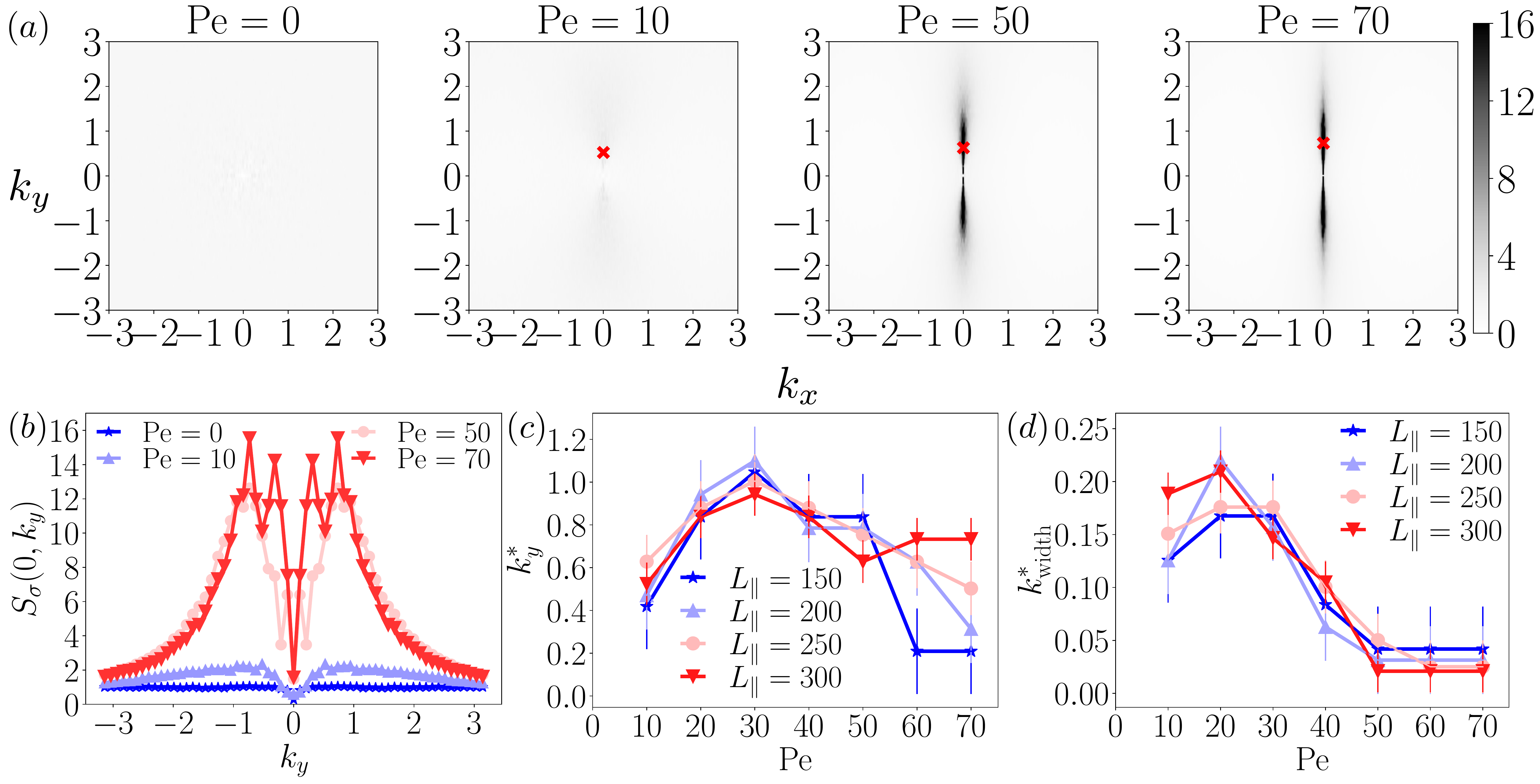}
    \caption{{(a)~$S_{\sigma}(\bm{k})$ at various $\rm{Pe}$ with $L_{\parallel} = 300$
    and $\phi=0.35$, corresponding to $N=8100$. The red arrows indicate the locations of the highest peaks. 
    (b)~$S_{\sigma}(0, k_y)$ at different $\rm{Pe}$ with $L_{\parallel} = 300$.  The peaks correspond to the $k^{*}_y$ in (c). 
    (c)~The characteristic wavevector $k^{*}_y$ extracted from $S_{\sigma}(\bm{k})$ against different $\rm{Pe}$. 
    (d)~The width of the lobes extracted from $S_{\sigma}(\bm{k})$ against different $\rm{Pe}$.   }
    }
    \label{fig:S-S_k_charge}
\end{figure*}

The finite-size scaling results in Fig.~\ref{fig:OP}(a) are \emph{not} consistent with any such transition.  For each system size, there is a crossover from homogeneous to de-mixed states on increasing $\Pe$.  However, increasing $N$ shows that ever larger values of $\Pe$ are required to see demixing: there is no sharp threshold $\Pe^*$ at which long-ranged order appears.  Figure~\ref{fig:OP}(a) shows that these results are independent of whether the system is initialised in a homogeneous (mixed) or fully demixed state.  

Figure~\ref{fig:OP}(a) suggests a correlation length which grows with $\Pe$ but never diverges. 
{We identify it with the width of the lanes (perpendicular to the drive), which is $L_\perp/2$ in state IV, but much smaller in state III, these correlations are discussed further below.
A similar scenario was proposed in~\cite{glanz2012nature} for correlations parallel to the drive; our focus here on perpendicular correlations is essential for detecting the fully de-mixed state \RN{4}, which resembles macroscopic phase separation at equilibrium.}
{Comparing with that case, our observation of}
%Compared with equilibrium phase separation, this result 
demixing in large finite systems without any phase transition is a distinctively non-equilibrium effect.

\subsection{Spatial correlations}

\subsubsection{Real-space analysis}

States \RN{2} and \RN{3} of Fig.~\ref{fig:Configuration} exhibit laning, but the particle density also develops inhomogeneities, where right- and left-moving particles tend to block each other.
To investigate these correlations, define density-density and ``charge-charge'' correlation functions:
\begin{equation}\label{eq:DensityCharge}
\begin{aligned}
  h_{\rho,\rho}(\bm{r}) & =\langle \rho(\bm{{r}})\rho(\bm{{0}}) \rangle/\rhobar^2 -  1 
  \\
  h_{\sigma,\sigma}(\bm{{r}}) & = \langle \sigma(\bm{{r}})\sigma(\bm{{0}}) \rangle / \rhobar^2 
\end{aligned}
\end{equation}
where $\rho(\bm{r}) = \sum_{i=1}^{N} \delta(\bm{r}_i -  \bm{r})$ is the empirical particle density, 
$\sigma(\bm{r}) = \sum_{i=1}^{N} \sigma_i \delta(\bm{r}_i -  \bm{r})$ is the ``charge'' density, and $\bar\rho=N/(L_\parallel L_\perp)$ is the mean density.

\begin{figure*}[t]
    \centering 
    \includegraphics[width=0.95\textwidth]{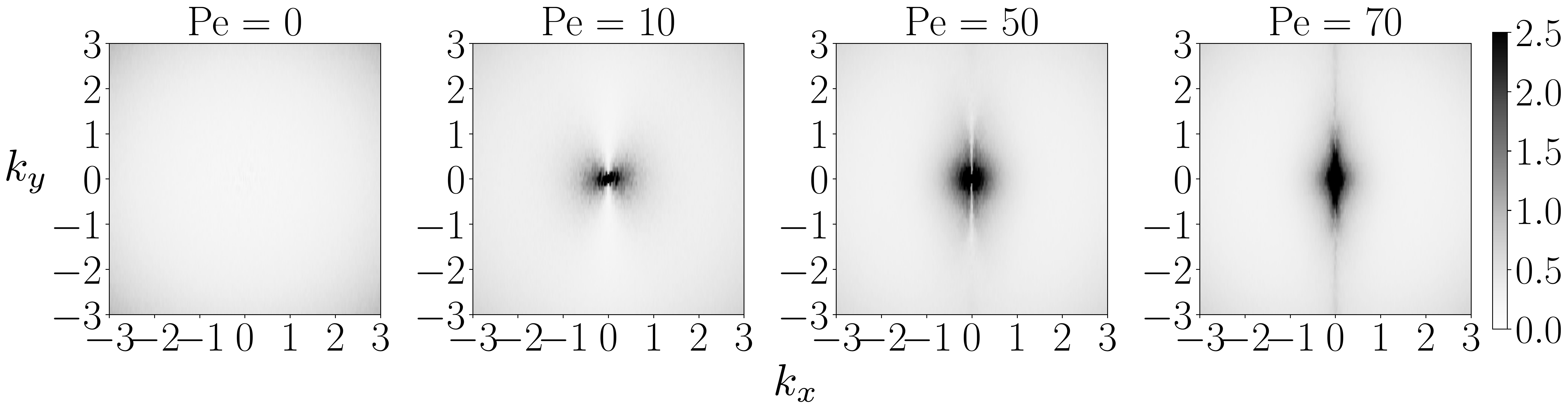}
    \caption{{$S_{\rho}(\bm{k})$ at various $\rm{Pe}$ with $L_{\parallel} = 300$ and $\phi = 0.35$, corresponding to $N = 8100$.  The features near the origin correspond to the emergence of TJCs. }
    }
    \label{fig:S-S_k_density}
\end{figure*}

The behaviour of these correlation functions on hydrodynamic length scales has been analysed~\cite{demery2016conductivity, poncet2017universal, geigenfeind2020superadiabatic, frusawa2022stochastic} within a mean-field approximation, which is valid
when density fluctuations are not too large, for example at high density with weakly interacting particles~\cite{poncet2017universal}.  For the system considered here, the particles have strongly repulsive cores and Fig.~\ref{fig:Configuration} shows that density fluctuations are large.  Hence mean-field theory is not expected to be quantitative: it does not predict the demixing effect seen in state \RN{4}.  Still, the universal hydrodynamic correlations of Refs.~\cite{poncet2017universal} should emerge on large scales, within homogeneous states.
Figure~\ref{fig:OP}(b) shows the charge autocorrelation function measured parallel to the drive.  For $\Pe=20$, {these results are consistent with Refs.~\cite{poncet2017universal}, which predicted a decay as  $|x|^{-3/2}$.} For larger Pe, the correlations grow smoothly with Pe, as the lanes become more pronounced.  

\subsubsection{Correlations in Fourier space}

{It is useful to display these correlations in Fourier space.  For the charge density $\sigma$,  the structure factor associated with two-point correlations is 
\begin{equation}
    S_{\sigma}(\bm{k}) = \frac{1}{N}\Big\langle \Big | \sum^{N}_{j=1} \sigma_j \exp(-i\bm{k}\cdot \bm{r}_j) \Big |^2\Big\rangle,
\end{equation}
where $\bm{k} = (k_{\parallel}, k_{\perp})$ is the wavevector.  
}

{Results are shown in Fig.~\ref{fig:S-S_k_charge}.
Figure~\ref{fig:S-S_k_charge}(a) shows that these functions have a characteristic shape, with strongly anisotropic correlations that were  already anticipated in Chapter 3.1 of Refs.~\cite{SCHMITTMANN1995alt2}, see also Sec.~\ref{sec:SF-MFT} below.  (The system is not phase separated for these parameters and system sizes.)
$S_\sigma$ has two lobes that lie along the $k_y$-axis, which get stronger as Pe increases.  To characterise these lobes, we consider $S_{\sigma}(0,k_y)$ [Fig. \ref{fig:S-S_k_charge}(b)] and we denote its maxima by $\pm k_y^*$.  

We identify $\pi/k_y^*$ as one half of the wavelength of the modulations in $\sigma(\bm{r})$, which is the typical lane width.
We also measure the width of the lobes in $S_\sigma$, by plotting $S_\sigma(k_x,k_y^*)$ as a function of $k_x$ and computing its full width at half maximum.  This is denoted by $k^*_{\rm width}$ and represents the range of correlations \emph{along the lanes}.

Figure~\ref{fig:S-S_k_charge}(c) shows how $k_y^*$ depends on Pe and system size (always at fixed density).  Similarly Fig. \ref{fig:S-S_k_charge}(d) shows $k^*_{\rm width}$.
Several points are notable: For small-to-moderate Pe, the length scale $\pi/k_y^*$ is a few particle diameters, independent of system size.  For the largest Pe and the smaller system sizes, one sees a change in behaviour of $k_y^*$, which corresponds to the onset of macroscopic phase separation.
At the smallest Pe, the correlations are weak but appear rather long-ranged, we return to this point in Sec.~\ref{sec:SF-MFT}, below.
 We also find that $k_{\rm width}^*$ is much smaller than $k_y^*$, consistent with the fact that correlations along the lane have a range that is much longer than the lane width.  For large Pe, the width starts to depend on the system size, which naturally limits the range of correlations (clearly, $k_{\rm width}^*\geq \pi/L_{\parallel}$).}

\begin{figure*}[t]
    \centering 
    \includegraphics[width=0.98\textwidth]{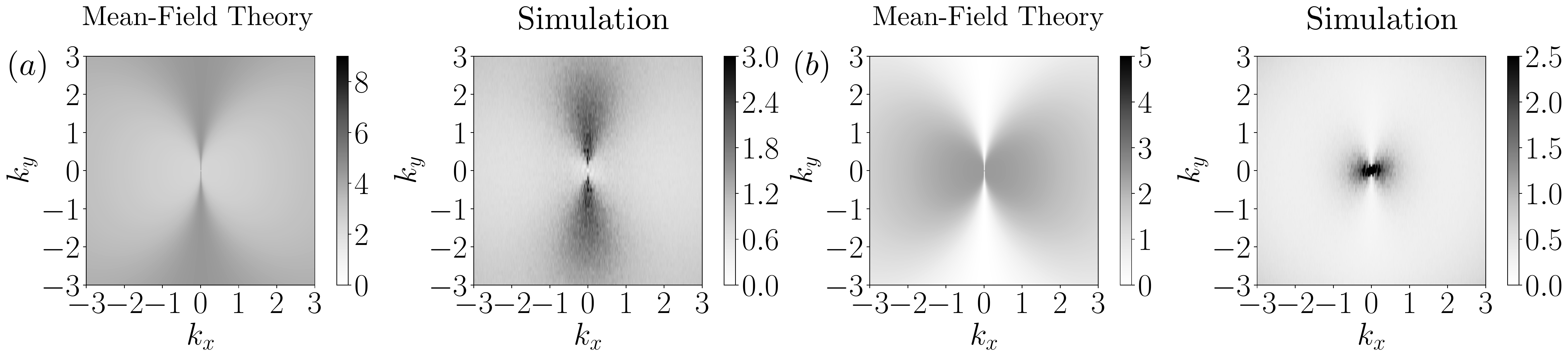}
    \caption{{(a)~$S_{\sigma}(\bm{k})$ from both mean-field theory and simulation. (b)~$S_{\rho}(\bm{k})$ from both mean-field theory and simulation.  Both $S_{\sigma}(\bm{k})$ and $S_{\rho}(\bm{k})$ are shown with $\rm{Pe} = 10$ and $\phi = 0.35$.  For simulations, $L_{\parallel} = 300$, corresponding to $N = 8100$.
    }}
    \label{fig:S-S_k_compare}
\end{figure*}

{To complement the analysis of charge correlations, we show the structure factor for the density field:}
\begin{equation}
    S_{\rho}(\bm{k}) = \frac{1}{N}\Big\langle\Big | \sum^{N}_{j=1} \exp(-i\bm{k}\cdot \bm{r}_j) \Big |^2\Big\rangle
    \; ,
\end{equation}
{which is proportional to the Fourier transform of $h_{\rho,\rho}$, see Fig.~\ref{fig:OP}(c).  
This function is shown in Fig.~\ref{fig:S-S_k_density}.  On increasing Pe, a significant peak appears at small $k$, signalling the formation of TJCs and associated large density fluctuations. 
(The compressibility $\tilde\chi$ can be expressed in terms of an orientational integral over $S_\rho$.)
However, this peak also has a significant orientation dependence, which is discussed in the next subsection by comparing with results of mean-field theory.}

\subsubsection{Comparison with mean-field theory}\label{sec:SF-MFT}

It is instructive to compare these results in Fourier space with those of mean-field theory~\cite{poncet2017universal}.
Based on Eq.~(9) from Refs.~\cite{poncet2017universal}, one can obtain the structure factors $S_{\sigma}(\bm{k})$ and $S_{\rho}(\bm{k})$:
\begin{equation}
\begin{aligned}
  S_{\sigma}(\bm{k}) &\sim  \frac{k^4+A_{\sigma} k^2_{\parallel}\rm{Pe}^2}{k^4+B_{\sigma} k^2_{\parallel}\rm{Pe}^2},
  \\
  S_{\rho}(\bm{k}) &\sim \frac{A_{\rho} k^2_{\parallel}\rm{Pe}^2}{k^4+B_{\rho} k^2_{\parallel}\rm{Pe}^2},
\end{aligned}
\label{equ:MF-Sk}
\end{equation}
where $A_{\sigma}, B_{\sigma}, A_{\rho}, B_{\rho}$ are positive constants that depend on the system of interest via a single parameter $\tilde v_0$, which is the Fourier transform of the interaction potential evaluated at wavevector $\bm{k} = 0$.   For the purposes of this paper, it only matters that these are positive constants, of order unity.

{Figure.~\ref{fig:S-S_k_compare} shows representative results from simulations above, compared with results of the theory, for the representative parameter value $\tilde v_0 = 1$, which leads to $A_{\sigma}, B_{\sigma}, A_{\rho}, B_{\rho}$ of order unity. The theory captures the anisotropic correlations at small $k$.  However it does not capture the peaks that appear in the structure factor, for example it does not make predictions for $k_y^*$, nor does it capture the peak in $S_\rho$ (which we attributed previously to TJCs).  An interesting prediction of Eqs.~\eqref{equ:MF-Sk} is that small Pe leads to weak correlations with large correlation lengths that scale as $\xi\sim1/({\rm Pe}\sqrt{C})$ where $C$ is one of $A_\sigma,A_\rho,B_\sigma,B_\rho$.  It would be interesting to investigate these length scales further, but the differences between theory and simulation are strong enough that we do not attempt this here.  (It is not clear to us if this theoretical prediction is linked to the increase of $k_y^*$ with Pe, observed for small Pe in Fig.~\ref{fig:S-S_k_density}.)}

\begin{figure*}[t]
    \centering
    \includegraphics[width=0.98\textwidth]{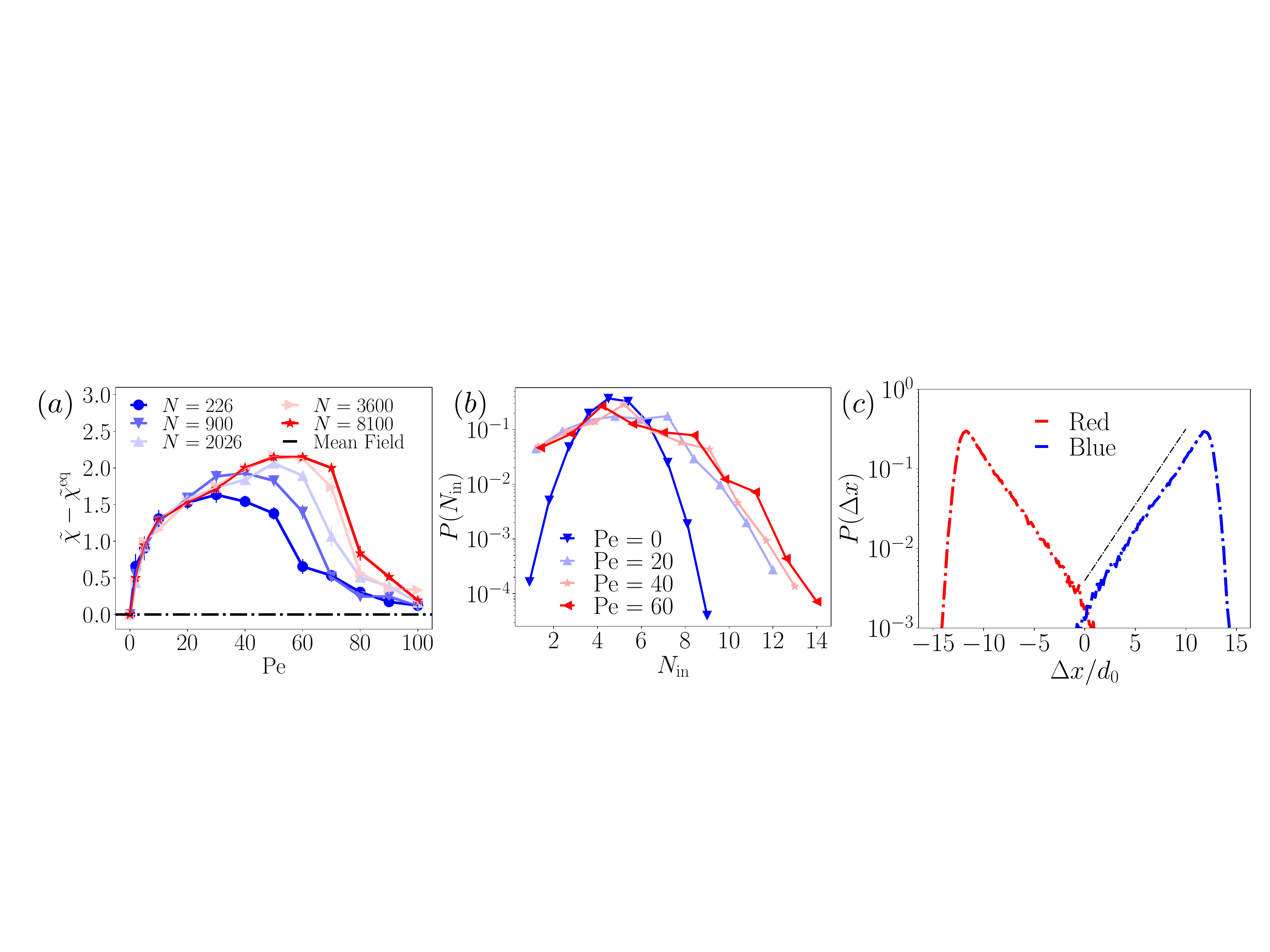}
    \caption{
    (a)~$\tilde{\chi}-\tilde{\chi}^{\text{eq}}$ against Pe, for different system sizes, with cut-off radius $R=5d_0$.  The mean-field theory predicts that this quantity vanishes for $R\to\infty$.  
    (b)~{Probability distribution for the number of particles in a randomly-chosen circular probe region of radius $r_0=1.75d_0$, for various $\text{Pe}$.}
    (c)~{Probability distributions for the $x$-displacement $\Delta x$ of each species of particles within time $\tau = 0.25\tau_B$, at $\rm{Pe} = 50$. }  The dashed line indicates $P(\Delta x) \propto e^{-(\Delta x_{\text{free}}-\Delta x)/\ell}$.  For (b), (c),  $L_\parallel=300d_0$.
    }
    \label{fig:compressibility}
\end{figure*}

\begin{figure}[t]
    \centering 
    \includegraphics[width=0.45\textwidth]{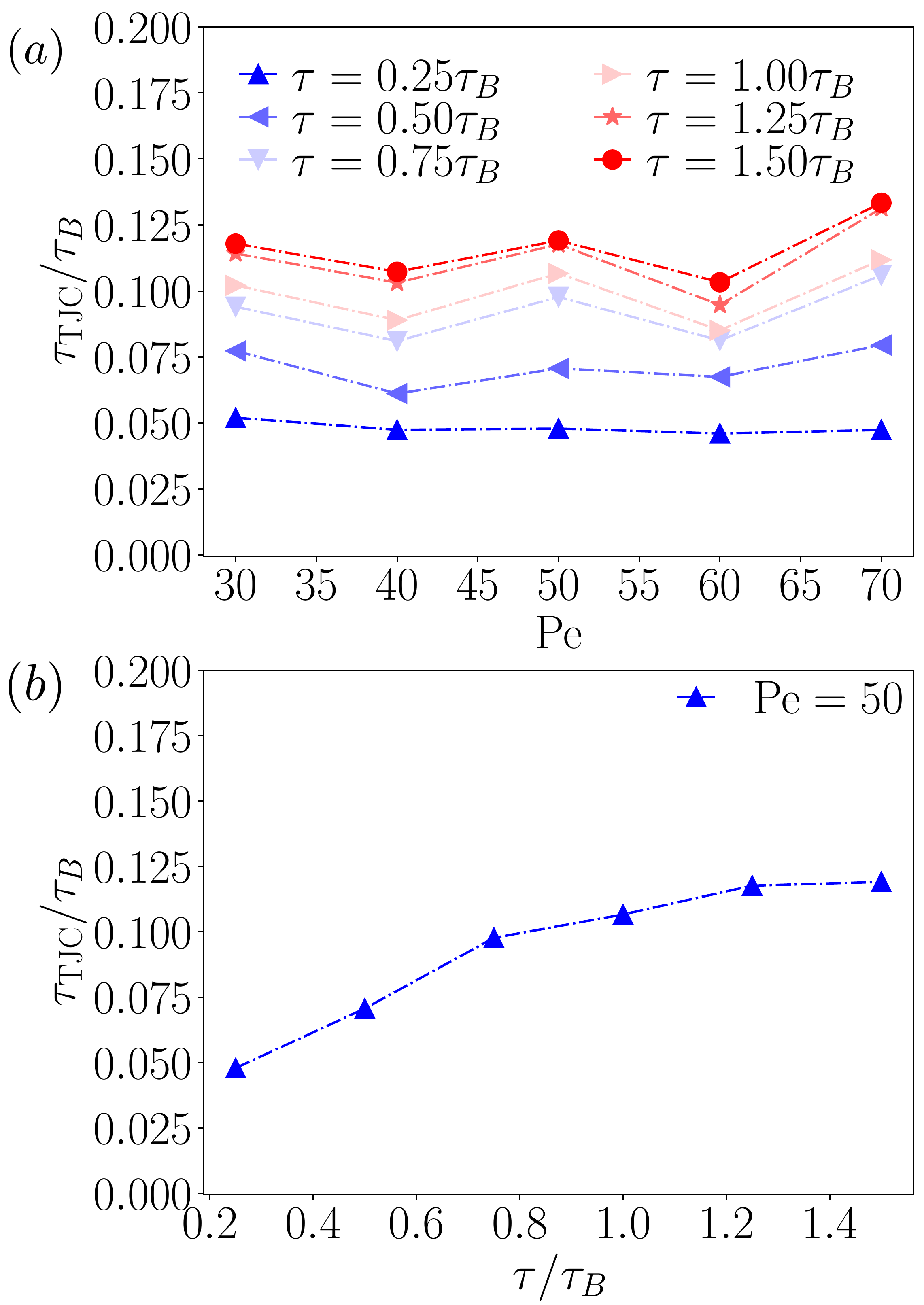}
    \caption{(a)~The characteristic trap time $\tau_{\rm TJC}$ in units of $\tau_B$ at different $\rm Pe$ number when TJCs are present with $L_{\parallel}=300d_0$.
    (b)~The characteristic trap time $\tau_{\rm TJC}$ against measurement window $\tau$ of system as in (a).  The characteristic trap time $\tau_{\rm TJC}$ converges with sufficiently large measurement window $\tau$.
    }
    \label{fig:S-Trap}
\end{figure}

\subsection{Transient jammed clusters}

{\subsubsection{Characterisation of TJCs}

States \RN{2} and \RN{3} in  Fig.~\ref{fig:Configuration} show strikingly large density fluctuations involving clusters of particles with opposite charge, which block each others' motion, leading to a transient jamming phenomenon.  
To analyse this, Fig.~\ref{fig:OP}(c) shows density correlations for $\Pe=(0,50,100)$,
corresponding to states \RN{1}, \RN{3}, and \RN{4} in Fig.~\ref{fig:Configuration}. For the equilibrium fluid state ($\Pe=0$), the correlations are isotropic and short-ranged.  For $\Pe=50$, the correlations are stronger (as indicated by the darker shading) and they are also longer ranged -- this is a signature of the TJCs. For $\Pe=100$, the correlations resemble those for $\Pe=0$.  This  re-entrant behaviour occurs because the demixed domains are similar to equilibrium systems that are being advected at a constant velocity.  {This advection also affects the usual nearest-neighbour correlations in $h_{\rho,\rho}$, leading to to bands in $h_{\rho,\rho}$, extended along the driving direction.}

To further quantify the TJCs, we integrate the correlation function over space to obtain an analog of the compressibility of an equilibrium system~\cite{hansen2013theory}: 
\beq\label{eq:compressibility}
\tilde{\chi}  = 
1 + \overline{\rho} \int_{|\bm{r}|< R} h_{\rho, \rho}(\bm{r}) d\bm{r} \; .
\eeq
The large-distance cutoff $R$ reduces statistical uncertainty, Appendix~\ref{app:chi} shows that these results depend weakly on $R$.

Figure~\ref{fig:compressibility}(a) shows $\tilde\chi$, measured relative to its equilibrium value.
It grows rapidly as Pe increases from zero, indicating the formation of TJCs.  There is a plateau at intermediate Pe, showing that these clusters persist across broad range of parameters.  At large Pe, $\tilde\chi$ decreases, due to the re-entrant demixing, recall Fig.~\ref{fig:OP}(c). 
The mean-field theory of Refs.~\cite{poncet2017universal} does not capture these TJCs, because of its  assumption of small density fluctuations. For example, the correlations predicted by that theory lead to $\tilde\chi\to0$ as $R\to\infty$ (see Appendix~\ref{app:chi}): this prediction is not consistent with the data presented here.

As an alternative characterisation of TJCs, Fig.~\ref{fig:compressibility}(b) shows {the probability distribution of the number of particles $N_{\rm in}$ in a randomly chosen circular probe region}~\cite{crooks1997gaussian, del2018energy, omar2021phase}.  
The variance of this distribution increases dramatically with $\Pe$, with tails at large $N_{\rm in}$ reflecting the existence of TJCs.  These clusters also have a dynamical signature: Fig.~\ref{fig:compressibility}(c)  shows the distribution of particle displacements parallel to the driving field.  Each particle displacement is measured as
\begin{equation}
    \Delta  x_{i}(\tau) = x_{i}(t_0+\tau) - x_{i}(t_0),
\end{equation}
where $t_0$ is an arbitrary initial time.  Then the results are averaged separately over the two species.  The distributions are skewed, with significant weight near zero displacement, reflecting particles which remain almost stationary in TJCs.   
Similar slowing down has been observed in $3$D~\cite{dutta2016anomalous, dutta2018transient, dutta2020length}. 

\begin{figure*}[t]
    \centering
    \includegraphics[width=0.98\textwidth]{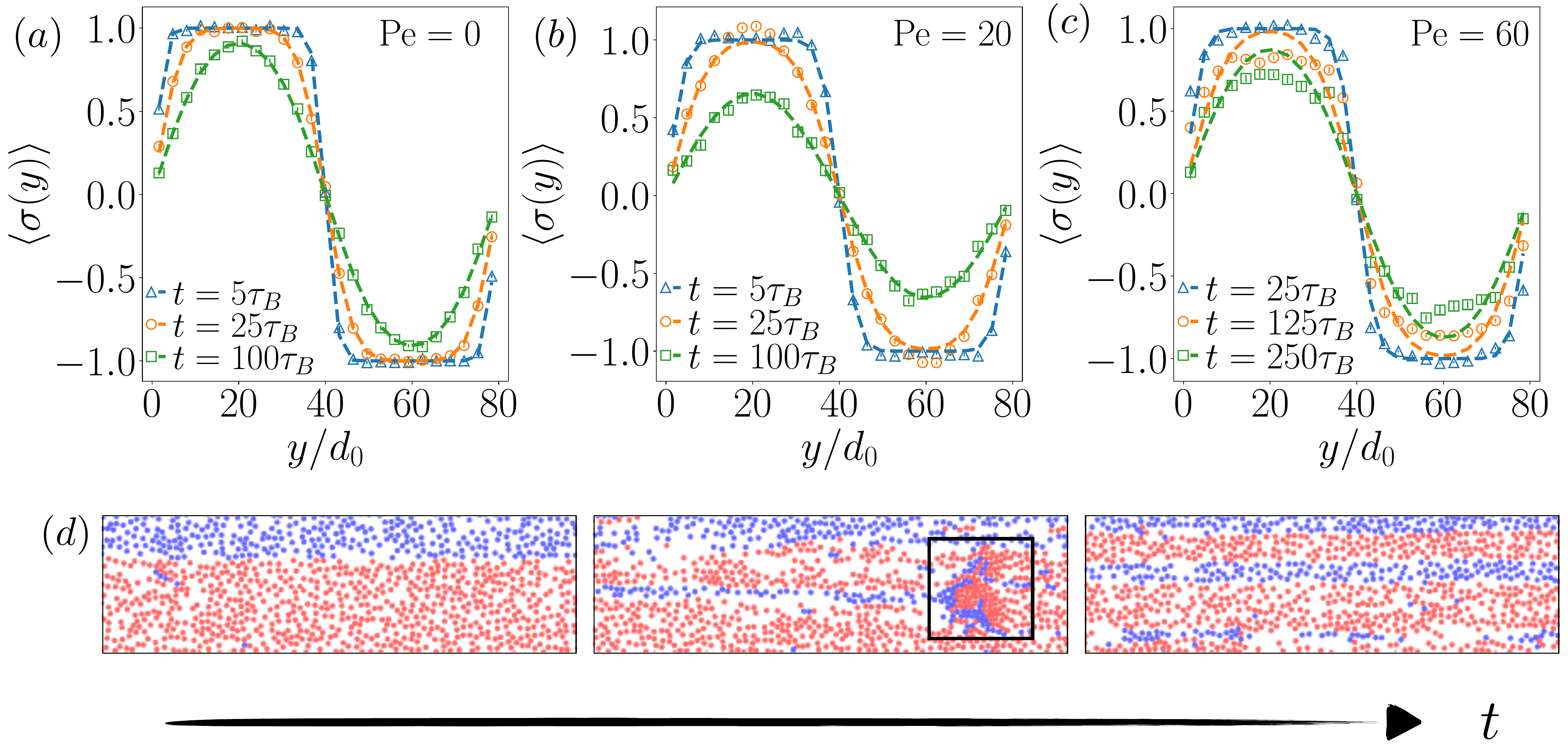}
    \caption{(a)-(c)~Time evolution of the charge density profile, starting from a demixed state, for $L_{\parallel}=100d_0$ and $S=1.25$.  {Symbols are numerical results and dotted lines are fits to the diffusion equation, see Appendix~\ref{app:interface} for details of the fitting procedure.} (d)~Time series, highlighting a TJC that nucleates new lanes.  We show a subsystem from  large simulation with $L_\parallel=300d_0$ and ${\rm Pe}=78$. 
    }
    \label{fig:interface}
\end{figure*}

\subsubsection{Residence time for particles in TJCs}
\label{sec:residence}

The characteristic trapping time of particles in TJCs can be estimated from the distribution of longitudinal displacement in Fig.~\ref{fig:compressibility}(c), using the following simple model.
The measurement window for the displacement is $\tau$ and we consider the species with positive charge, so a
free particle (outside of the TJC) travels an average distance approximately $\Delta x_{\text{free}}=\tau E/\gamma$ within this time. (The average velocity of a free particle is $v_{\text{free}} = E/\gamma$.)  

From Fig.~\ref{fig:compressibility}(c), the distribution of displacements $P(\Delta x)$ has an exponential tail for small $\Delta x$, we write 
\beq
P(\Delta x)\approx A e^{-(\Delta x_{\text{free}}-\Delta x)/ \ell}
\label{equ:Pdx}
\eeq
where $A$ is a constant, $\ell$ is a characteristic length scale, which is an estimate of the length by which a typical particle in a TJC is ``held back'' by its temporary immobilisation.  (The value of $A$ presumably depends on how many particles participate in TJCs.)  As a simple method to obtain a characteristic trapping time we divide $\ell$ by the characteristic velocity to obtain $\tau_{\rm TJC} = \ell/v_{\text{free}}$.  

To better interpret this time scale, suppose that particles in TJCs stay there for an exponentially distributed time $t_{\rm TJC}$ with mean $\tau_{\rm TJC}$, during which time they do not move at all parallel to the field.  Their resulting displacement is then $\Delta x_{\rm free} - v_{\text{free}} t_{\rm TJC}$.  Assuming also that $P(t_{\rm TJC}) \propto {\rm e}^{-t_{\rm TJC}/ \tau_{\rm TJC}}$, the tail of the displacement distribution is readily seen to be Eq.~\eqref{equ:Pdx} with $\ell = \tau_{\rm TJC} v_{\text{free}}=\tau_{\rm TJC} d_0{\rm Pe}/\tau_{\rm B}$. 

However, this simple analysis neglects effects of the measurement time window $\tau$.  In particular, the distribution of $\Delta x$ is not sensitive to trapping times longer than $\tau$.  As a result, the above argument tends to underestimate the trapping time.
Figure~\ref{fig:S-Trap}(a) shows the resulting $\tau_{\rm TJC}$ in the regime $\rm Pe \in [30, 70]$.  These results indicate that large measurement times $\tau$ can be used to obtain robust estimates for the trapping time.  Specifically, taking $\text{Pe}= 50$ and $\tau = \tau_B$, we obtained 
$\ell = 5.4d_0$ from the exponential fit and therefore have $\tau_{\rm TJC} = 0.11\tau_B$.  While this is smaller than the Brownian time, it is significantly larger than the the time for a particle in a lane to move its own diameter, which is $\tau_B/{\rm Pe}\approx 0.020\tau_B$.
However, smaller $\tau$ results in underestimates of the true $\tau_{\rm TJC}$, as expected, see Fig.~\ref{fig:S-Trap}(b).  Nevertheless, the resulting estimates of $\tau_{\rm TJC}$ are all of a similar order of magnitude; they also depend weakly on Pe, within the range $30<{\rm Pe}<70$ where TJCs play a significant role.

\subsection{Dynamics of mixing -- the role of TJCs} 

We have shown that these driven systems are always mixed in the limit of large systems and that the mixed state supports large TJCs.  We now establish a connection between these two results.  

We choose parameters such that the system's steady state is mixed but we initialise it in a demixed state, so that we can follow the mixing process.
To collect good statistics while maintaining a sufficiently large $L_\perp$, we reduce the aspect ratio to $S=1.25$.
Figure~\ref{fig:interface} shows time-dependent composition profiles perpendicular to the drive, for various Pe. 
In the absence of driving ($\Pe=0$), mixing happens by diffusion.  To illustrate this, we fit the results to the (analytical) solution of the diffusion equation $\partial_{t} \langle\sigma\rangle = D_{y} \partial_{y}^2 \langle\sigma\rangle$, using $D_{y}$ as a single fitting parameter (see Appendix~\ref{app:interface} for details of the fitting procedure).
The fit is excellent.  For $\Pe=20$, the system still mixes readily: the diffusion equation still fits the data adequately but the value of $D_{y}$ is larger, reflecting the ELD~\cite{klymko2016microscopic, yu2022lane}. 

However, for $\Pe=60$,  the lifetime of the demixed state is longer: the interface between the two domains is stable on short time scales, because collisions between oppositely charged particles tend to deflect them back into their own domains.   On longer time scales, mixing does occur, but the mechanism is not diffusive, as signalled by systematic deviations from the diffusive theory in Fig.~\ref{fig:interface}(c).  {In particular, the composition inside the demixed domains relaxes more quickly than diffusive mixing would predict, indicating that a collective mechanism is disrupting the domains.}

This mixing mechanism is illustrated in Fig.~\ref{fig:interface}(d) and the Supplemental Material movies 4,5~\cite{SM}:  collisions near the interface occasionally form TJCs which disrupt the flow and mix the counterpropagating species.  This initiates a period of chaotic flow, which involves further TJCs and -- eventually -- the creation of new stable lanes.
This mechanism further emphasizes the rich phenomenology of these non-equilibrium systems.

To understand the absence of demixed states in large systems, observe that increasing $L_\parallel$ provides more locations where TJCs can form, initiating the instability.  This mechanism is intrinsic to the interface between the oppositely-moving domains: Appendix~\ref{app:interface} shows that increasing $L_\perp$ (at fixed $L_\parallel$) does not change the local behaviour near the interface.  
In other words, the absence of de-mixed states for large $L_\parallel$ depends weakly on the aspect ratio of the system, since any interface between counterpropagating domains is prone to fragmentation via TJCs. 
Interfaces between oppositely-driven domains have been recently studied in Refs.~\cite{del2019interface, dean2020effect}, and nucleation of lanes is also reminiscent of other complex flow phenomena~\cite{bouchet2019rare}.

\begin{figure}[t]
    \centering 
    \includegraphics[width=0.49\textwidth]{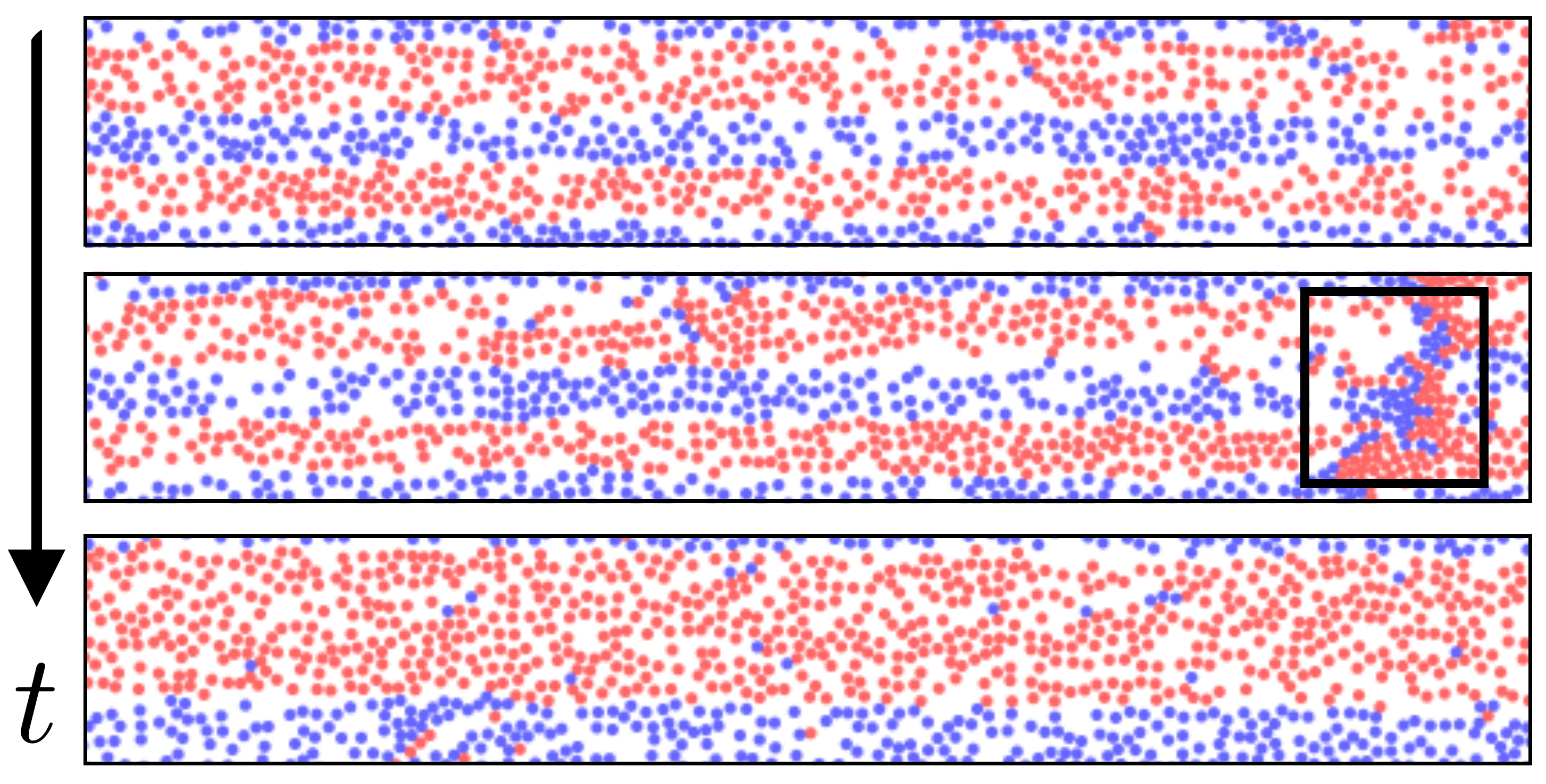}
    \caption{
    Time series, highlighting a TJC that causes coarsening of two lanes.  We show a subsystem from  large simulation with $L_\parallel=150d_0$ and ${\rm Pe}=80$.
    }
    \label{fig:S-Coarsen}
\end{figure}

If the system is small enough that the steady state is fully demixed, the mechanism of Fig.~\ref{fig:interface}(d) can also operate in reverse: four lanes can merge into two,  via TJC formation at the interface, see Fig.~\ref{fig:S-Coarsen}.  However, for large systems, the mixing (lane creation) events  predominate.

\section{Outlook} 
 Lane formation for oppositely driven particles relies on ELD, which tends to cause demixing, perpendicular to the drive.  We showed that this demixing is not a phase transition -- sufficiently large systems always remain mixed.  They also support large density fluctuations in the form of TJCs, which exist over a wide range of Pe, and play an important role in mixing, by destabilising interfaces between domains with oppositely driven particles.  
In this sense, the complex patterns shown in Fig.~\ref{fig:Configuration} reflect a competition between ordering by ELD and mixing via TJCs.  We look forward to future work on such patterns. For example, the possibility of system-spanning TJCs may be relevant for dynamical arrest in small systems~\cite{helbing2000freezing}.  {Similar phenomena may also be relevant for pattern formation in other non-equilibrium systems, such as self-propelled particles or sheared fluids~\cite{wensink2012emergent,Farrell2012,bain2017critical,del2018energy,Bar2020,Stopper2018, solon2022susceptibility,anderson2023ant,zhang2022arxiv,keta2023arxiv}, and the control of such patterns offers possibilities for the design of directed self-assembly~\cite{chennakesavalu2021probing, vansaders2023informational}.  At higher densities, we also anticipate an interesting interplay between TJCs and crowding (jamming or glassy) behaviour~\cite{berthier2019glassy,keta2022disordered}.  
}

\begin{acknowledgments}
We thank K. Thijssen, D. Frenkel, L. Berthier, O. B{\'e}nichou, M. Cates,  and N. Wilding for helpful discussions.
\end{acknowledgments}

\begin{appendix}

\section{Model and simulation details}
\label{app:model}

We non-dimensionalise the Langevin Eq. (1) using base units $d_0,m,\epsilon$.  The natural unit of time is $t_0 = d_0\sqrt{m/\epsilon}$.  Then we define reduced quantities which are: the particle position $\tilde{\bm{r}}_i = \bm{r}_i/d_0$, the Peclet number $\text{Pe}= E d_0/(k_{\rm B} T)$, the volume fraction $\phi=N\pi d_0^2/(4 L_\parallel  L_\perp)$, the reduced time $\tilde t=t/t_0$,  the reduced interaction energy $\tilde U=U/\epsilon$, the reduced friction constant $\tilde\gamma=\gamma t_0/m$,  the reduced temperature $\tilde T=k_{\rm B}T/\epsilon$, and the aspect ratio of the simulation box $L_\parallel/L_\perp$. 
Recall that the Brownian time is $\tau_{\rm B} = d_0^2\gamma/(k_BT)=t_0\tilde\gamma/\tilde{T}$: this evaluates to $10^3 t_0$ for the parameters used in this paper.

\begin{figure*}[t]
    \centering 
    \includegraphics[width=0.98\textwidth]{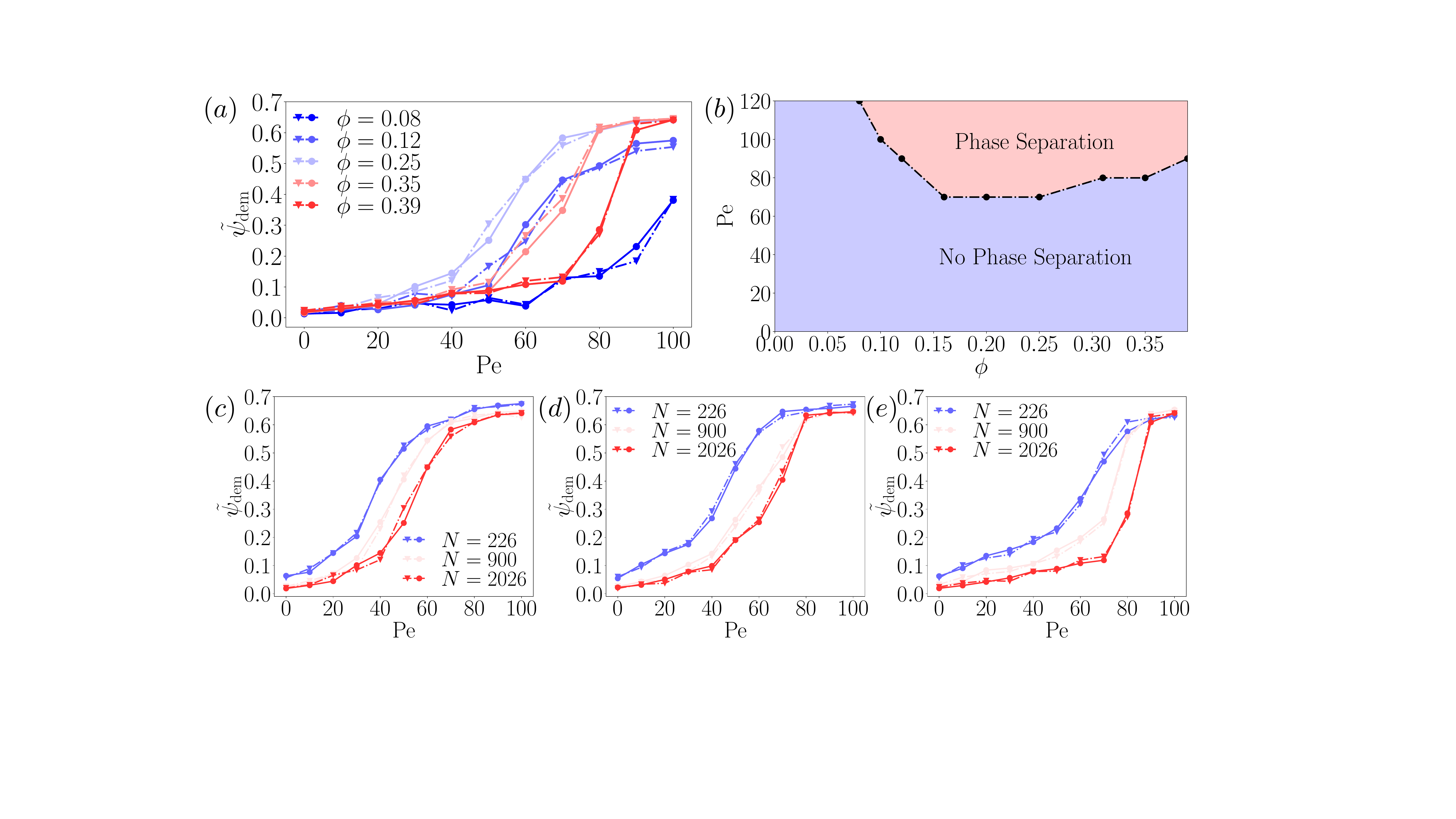}
    \caption{(a)~$\tilde{\psi}_{\rm{dem}}$ against $\rm{Pe}$ at different packing fraction $\phi$ for $N = 2026$.  
    (b)~Diagram showing the regime in which phase separation is observed ($\tilde{\psi}_{\rm{dem}}\gtrsim 0.6$) for $N=2026$.  
    (c)-(e)~$\tilde{\psi}_{\rm{dem}}$ against $\rm{Pe}$ at different system sizes for $\phi = 0.25, 0.31, 0.39$ respectively. }
    \label{fig:S-OP_density}
\end{figure*}

The non-dimensionalised Langevin equation is then 
\beq
\frac{d^2}{d\tilde{t}^2}\tilde{\bm{r}}_{i}= - \tilde{\gamma} \frac{d}{d\tilde{t}}\tilde{\bm{r}}_{i} - \tilde{\nabla}_{i} \tilde{U}
+  \sigma_i  \textrm{Pe}\,\tilde{T} \hat{\bm{x}}  + \sqrt{2\tilde{\gamma} \tilde{T} }\, \tilde{\bm{\eta}}_i(\tilde{t}),
\eeq
where $\tilde{\bm{\eta}_i}$ is non-dimensionalised Gaussian white noise with mean zero and 
$\langle \tilde{{\eta}}_{i,\alpha}(\tilde{t}) \tilde{{\eta}}_{j,\beta}(\tilde{t}') \rangle = \delta_{ij}\delta_{\alpha\beta}\delta(\tilde{t}-\tilde{t}')$.

%\subsection{Simulation Details}
Simulations are performed in LAMMPS \cite{LAMMPS} with time step $\delta t = 5\times 10 ^{-4}t_0$.  
For initialisation, we either use disordered or demixed initial conditions.  For the disordered case, we initialise $N/2$ red and $N/2$ blue particles with random positions and simulate the equilibrium system ($E=0$) for $10^2\tau_B$, after which the force $E$ is introduced.  For the demixed case, we initialise $N$ particles of a single type with random positions and simulate the equilibrium system for $10^2\tau_B$.  Then we assign types to particles according to their $y$-coordinates: for  $y_i \in [0, L_{\perp}/2]$ we take $\sigma_i=-1$ and for $ y_i \in [L_{\perp}/2, L_{\perp}]$ we take $\sigma_i=1$.  Then we turn on the force $E$. 
As discussed in the main text, the steady state of the system is independent of whether disordered or demixed initialisation is used, and all measurements are taken in the steady state.

For Figs.~\ref{fig:Configuration}-\ref{fig:S-Trap} and \ref{fig:S-OP_density}-\ref{fig:S-Correlation} the steady state measurements are performed after simulation for at least $10^3 \tau_B$, so that the system has relaxed to its steady state.  We confirm the steady state behavior by comparing the simulations performed with both disordered and demixed initial conditions.  When the system phase separate, the coarsening process is extremely slow as the interfaces are smooth and stable.  When $\text{Pe} > 90$, it takes approximately $4\times 10^3 \tau_B$ for the system to reach complete phase separation, and results are collected after this time.

\section{Density dependence of $\psi_{\mathrm{dem}}$}
\label{app:density}

We defined $\psi_{\rm{dem}}$ to characterise the steady state behaviour of the system, perpendicular to the drive.  In this section, we investigate the density dependence of the steady state behaviour.  We do this by fixing the number of particles $N$ and the aspect ratio of the system $S$, while varying the system size.  We re-scale the order parameter so that results are comparable between systems at different densities, specifically
\begin{equation}
    \tilde{\psi}_{\rm{dem}} = \frac{L_{\parallel}L_{\perp}}{N} \psi_{\rm{dem}} \; .
\end{equation}
To justify this choice, note that a perfectly phase separated state has ``charge'' density $\sigma(\bm{r}) = + \overline{\rho}$ when $y\in[0, L_{\perp}/2)$, and $\sigma(\bm{r}) = - \overline{\rho}$ when $y\in[L_{\perp}/2, L_\perp)$, one has
\begin{equation}
   \tilde{\psi}_{\rm{dem}} \approx  \Big\langle\Big|\frac{1}{N} \int \sigma(\bm{r}) e^{-i\bm{k}^*\cdot \bm{r}} d\bm{r}\Big|\Big\rangle = \frac{2}{\pi} \approx 0.64,
\end{equation}
which is independent of the total density.

Simulation results are shown in Fig.~\ref{fig:S-OP_density}.  Panels (a) and (b) show data for a system of $N=2026$ particles.  We identify phase-separated states as those with $\tilde{\psi}_{\rm{dem}}> 0.6$, and we determine the phase boundary (black line) in Fig.~\ref{fig:S-OP_density}(b) by identifying $\rm Pe$ closest to $\tilde{\psi}_{\rm{dem}}= 0.6$ .  We emphasize that this ``phase boundary'' was computed by varying the box size with $N = 2026$; results will be different for systems with different numbers of particles.  
One sees that phase separation occurs 
across a range of concentrations $\phi$.  At low concentrations $\phi\lesssim 0.05$, particle collisions are rarer, which reduces the ELD effect, and suppresses laning and phase separation.   The tendency towards laning is non-monotonic in $\phi$ is consistent with previous work~\cite{chakrabarti2004reentrance}.
Figures~\ref{fig:S-OP_density}(c)-~\ref{fig:S-OP_density}(e) show finite-size scaling results for the order parameter, at three different densities within the laning regime.  In all cases, the Pe required for phase separation increases with $N$, indicating that sufficiently large systems will always remain mixed, for any given Pe.

\begin{figure*}[t]
    \centering 
    \includegraphics[width=0.98\textwidth]{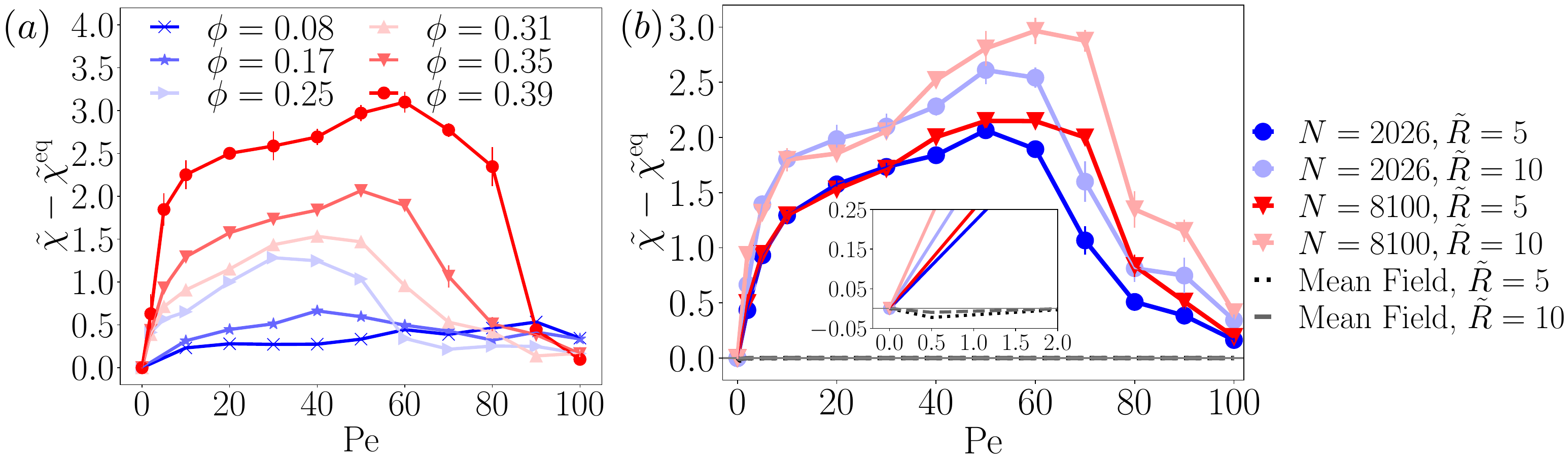}
    \caption{{(a)~Density fluctuation $\tilde{\chi}-\tilde{\chi}^{\text{eq}}$ against Pe at $N=2026$ with $\tilde R = 5$ for various packing fractions.} (b)~Density fluctuation $\tilde{\chi}-\tilde{\chi}^{\text{eq}}$ against Pe at $L_{\parallel}=150d_0$, $L_{\parallel}=300d_0$ with $\tilde R = 5, 10$.  The mean-field curve is shown with $\tilde{v}_0=1$ (see Appendix~\ref{sec:MFT}).
    }
    \label{fig:S-Compressibility}
\end{figure*}

For larger concentrations  $\phi>0.4$, we observe different qualitative behaviour, where neither laning nor phase separation takes place.  Instead the system forms macroscopic domains of the two species whose sizes and shapes fluctuate in time, as they flow past each other.  This regime brings even further computational challenges associated with the time required to converge to a steady state, and the ability to control finite size effects, so we do not address it here.  (These aspects were already challenging for the results presented here at lower density, but our extensive simulations have allowed us to control them.)

\section{Generalised compressibility $\tilde \chi$}
\label{app:chi}

\subsection{Dependence on density}
{In the main text, we showed the behavior of generalised compressibility $\tilde\chi$ for $\phi = 0.35$.  In Fig.~\ref{fig:S-Compressibility}(a) we show in the laning regime the behavior of $\tilde\chi$ for different packing fraction is the same.  It grows at low $\rm{Pe}$ due to the formation of TJCs and it plateaus at intermediate $\rm{Pe}$, corresponding to the coexsitence of TJCs and lanes at intermediate $\rm{Pe}$. At large $\rm{Pe}$,  $\tilde\chi$ drops at large $\rm{Pe}$, corresponding to the demixing of the particles and disappearance of TJCs.

From Fig.~\ref{fig:S-Compressibility}(a), we also observe that higher packing fraction corresponds to higher $\tilde\chi$ across all $\rm{Pe}$ regimes when TJCs are present.  The behavior is intuitive: at higher packing fractions particles become more packed when moving in opposite directions and form larger TJCs.

We also note the drop in $\tilde\chi$ at high $\rm{Pe}$ corresponding to demixing is density dependent: higher packing fraction requires higher $\rm{Pe}$ to demix.  The behavior is consistent with the density dependence of the order parameter as shown in Fig.~\ref{fig:S-OP_density}(a).
}

\subsection{Dependence on cutoff $R$}

In the main text, we defined $\tilde\chi$ in terms of the density fluctuations of the system.  In equilibrium systems, this $\tilde\chi$ is related to the compressibility (defined in terms of dependence of the volume on the applied pressure).  In these non-equilibrium systems there is no relation between $\tilde\chi$ and responses to applied pressure, so we refer to it as a generalised compressibility.

To parameterise the cutoff in Eq.~(\ref{eq:compressibility}), we non-dimensionalise as $\tilde R = R/d_0$.  
For the main text, we take $\tilde R=5$ to reduce numerical uncertainties.  In Fig.~\ref{fig:S-Compressibility}(b), we show that the behavior of $\tilde \chi$ depends weakly on this cutoff.  There is a small increase on increasing $\tilde R$, from the tails of $h_{\rho,\rho}$ but the robust qualitative feature is that $\tilde\chi$ increases strongly with $\rm Pe$, saturates at a plateau, and then decreases.

\subsection{Mean-field analysis}\label{sec:MFT}

We describe here the mean-field prediction for density-density correlations, see also Fig.~\ref{fig:compressibility}(a) of the main text.  From Ref.~\cite{poncet2017universal}, one obtains the hydrodynamic (or singular) part of the density-density correlation function which we denote as $h^s_{\rho,\rho}(x, y)$.
This hydrodynamic correlation is valid on large length scales but it misses effects of particle packing, for example one has $h^s=0$ for the equilibrium case $\mathrm{Pe}=0$ (although of course $h\neq0$ in this case).  Hence we subtract the equilibrium part of $\tilde\chi$ when comparing the numerical results with the theory.

The singular part of the correlation function in two spatial dimensions can be obtained from Ref.~\cite{poncet2017universal} as
\begin{equation}
    h^s_{\rho,\rho}(x, y)= \mathrm{Pe}^{1/2} \frac{H_{\rho,\rho}}{|x|^{3/2}} 
    %g
    \left(
    \frac{\text{Pe}\, y^2}{{\cal D}|x|} - 1\right)
    \exp\left( -\frac{\mathrm{Pe}\, y^2}{{2\cal D}|x|}\right),
\label{equ:h-mft}
\end{equation}
where $H_{\rho,\rho}$ and ${\cal D}$ are constants that are fully determined by the parameter $\tilde v_0$  that was discussed in Sec.~\ref{sec:SF-MFT}.
We always have ${\cal D}>0$; for $\tilde v_0 \lesssim 1.3$ then $H_{\rho,\rho}>0$ while for larger $\tilde v_0$ we have $H_{\rho,\rho}<0$.  

We integrate the singular part of the density correlation function to get the singular part of the generalised compressibility, $\tilde \chi^s_{\text{mf}}=\overline{\rho} \int_{|{\bm{r}}|< R} h^s_{\rho, \rho}({\bm{r}}) d{\bm{r}}$.  We obtain
\begin{multline}\label{eq:MFTchi}
    \tilde{\chi}^s_{\text{mf}} 
    = \overline{\rho}H_{\rho, \rho}\sqrt{\text{Pe}}
    \int^{R}_{0}\int^{\sqrt{R^2- x^2}}_{0} \bigg[ \frac{1}{ x^{3/2}}\left(\frac{\text{Pe}\, y^2}{2{\cal D} x} - 1 \right)
      \\  \times
      \exp\left({-\frac{\text{Pe}\, y^2}{2{\cal D} x}}\right) \bigg] d y d x \; . 
\end{multline}
Integrate $y$ by parts, and make the change of variable $u = x/R$ to simply the integral:
\begin{multline}\label{eq:MFTintegral}
    \tilde{\chi}^s_{\text{mf}} 
     =-\overline{\rho}H_{\rho, \rho}\sqrt{\text{Pe}/R} \\ \times
    \int^{1}_{0} \frac{\sqrt{1-u^2}}{u^{3/2}} \exp\left[{\frac{\text{Pe}\, R }{2{\cal D}} \left(u-\frac{1}{u}\right)}\right] du
    \; .
  \end{multline}
This final integral is easily computed numerically.  Moreover, the argument of the exponential is always negative so one easily sees that $\tilde{\chi}^s_{\text{mf}} \to0$ as the cutoff $R\to\infty$.  Figure~\ref{fig:S-Compressibility}(b) shows the $R$-dependence of results from simulation, showing that $\tilde\chi$ increases weakly with $R$, contrary to this prediction.  (Agreement is not expected here, because the mean-field theory does not capture large density fluctuations such as TJCs.)

\begin{figure}
    \centering 
    \includegraphics[width=0.45\textwidth]{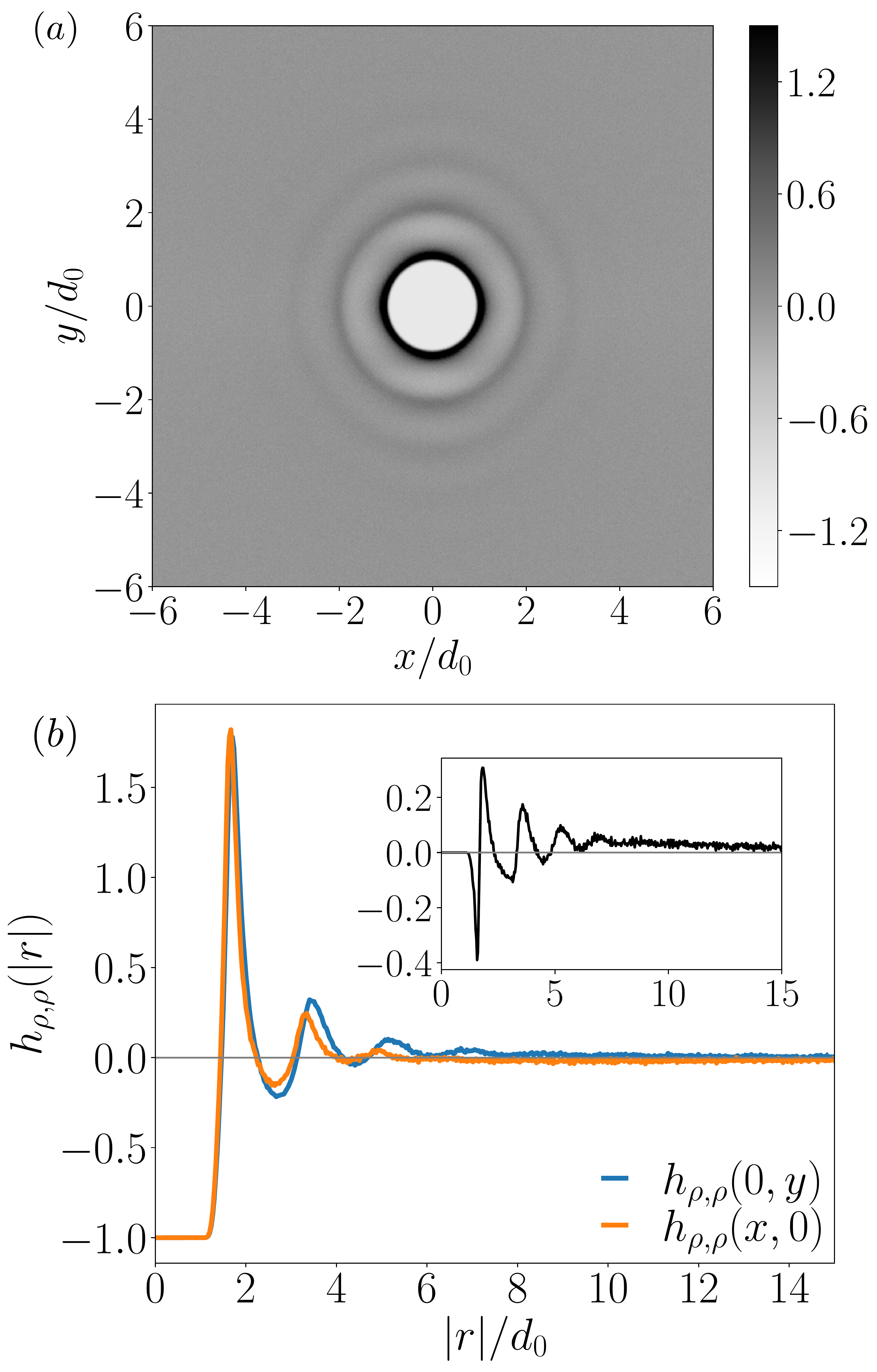}
    \caption{(a)~Dependence of $h_{\rho,\rho}(\bm{r})$ on $\bm{r}=(x,y)$. 
    (b)~The density-density correlation function $h_{\rho, \rho}(0, y)$ and $h_{\rho, \rho}(x, 0)$. The inset is the difference $h_{\rho, \rho}(0, y) - h_{\rho, \rho}(x, 0)$.  The sign of $h_{\rho, \rho}(0, y) - h_{\rho, \rho}(x, 0)$ suggests $H_{\rho, \rho} > 0$ in the mean-field theory.  For both (a) and (b), $L_\parallel=150d_0$ and $\rm Pe = 20$.
    }
    \label{fig:S-Correlation}
\end{figure}

For a more detailed analysis, Fig.~\ref{fig:S-Correlation} shows the behaviour of $h_{\rho,\rho}$ for ${\rm Pe}=20$.  In particular, we show the behaviour along the axes $\bm{r}=(x,0)$ and $\bm{r}=(0,y)$. 
For the latter case, Eq.~(\ref{equ:h-mft}) predicts that the hydrodynamic part of the correlation is exponentially small but the numerical results show a small but positive contribution that decays slowly at large scales.  (We attribute this to TJCs that are extended along the $y$-direction, which are not accounted for by mean-field theory.)
Since $h_{\rho,\rho}(x,0)<0$ at large distances while $h_{\rho,\rho}(0,y)>0$, the best chance for agreement with Eq.~\eqref{equ:h-mft} is to take $H_{\rho,\rho}>0$, indicating $\tilde{v}_0<1.3$.  The mean-field results are obtained with $\tilde{v}_0=1$, however, this leads to $\tilde{\chi}^s_{\text{mf}}<0$, see Fig.~\ref{fig:S-Compressibility}(b) (recall also that this $\tilde{\chi}^s_{\text{mf}}\to0$ for large $R$).

\section{Evolution of Interface}
\label{app:interface}

In this section we describe how we fit the time-evolution of the interface within demixed state described in the main text.  We provide evidence that the non-diffusive behavior of the time-evolution of the interface at the high $\text{Pe}$ regime depends weakly on the shape of the simulation box.  

\begin{figure*}[t]
    \centering 
    \includegraphics[width=0.95\textwidth]{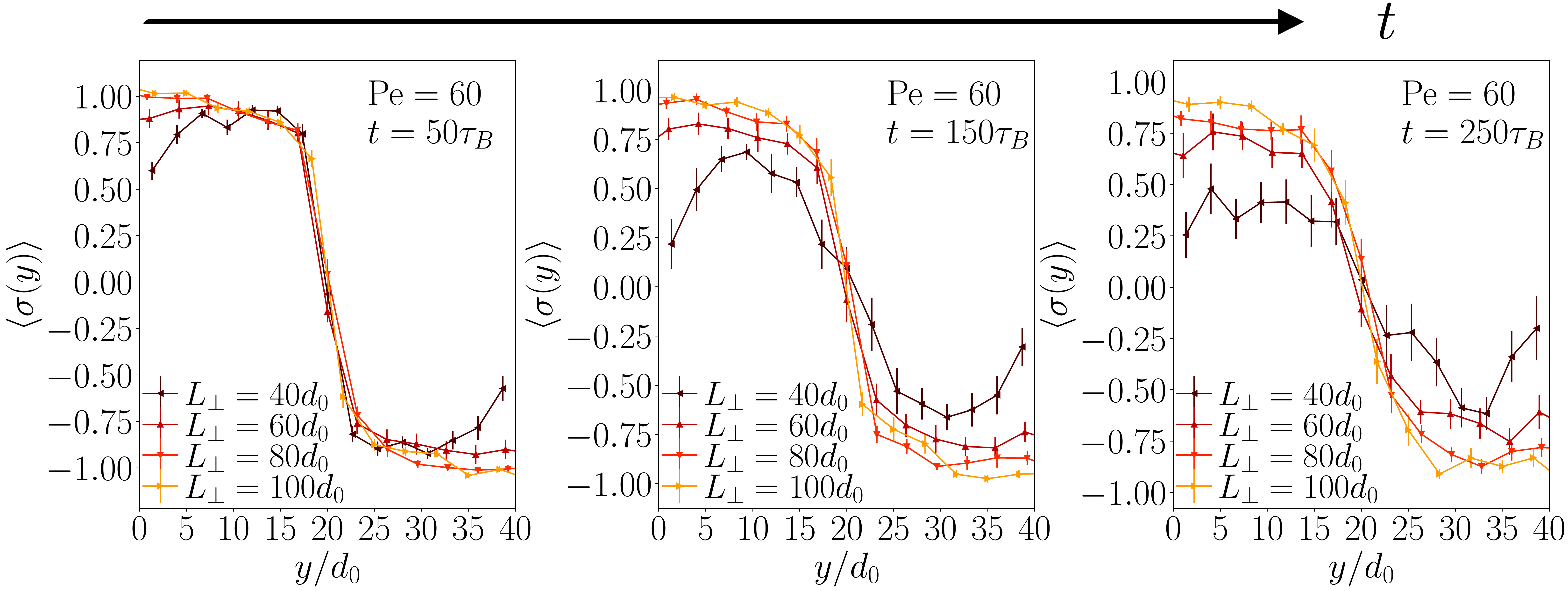}
    \caption{The charge density profiles  for systems with different aspect ratios near the interface at $t = 25\tau_B, 75\tau_B, 125\tau_B$ for $\text{Pe}=60$ with $L_{\parallel} = 100d_0$.
    }
    \label{fig:S-Interface-time}
\end{figure*}

\begin{figure*}
    \centering 
    \includegraphics[width=0.95\textwidth]{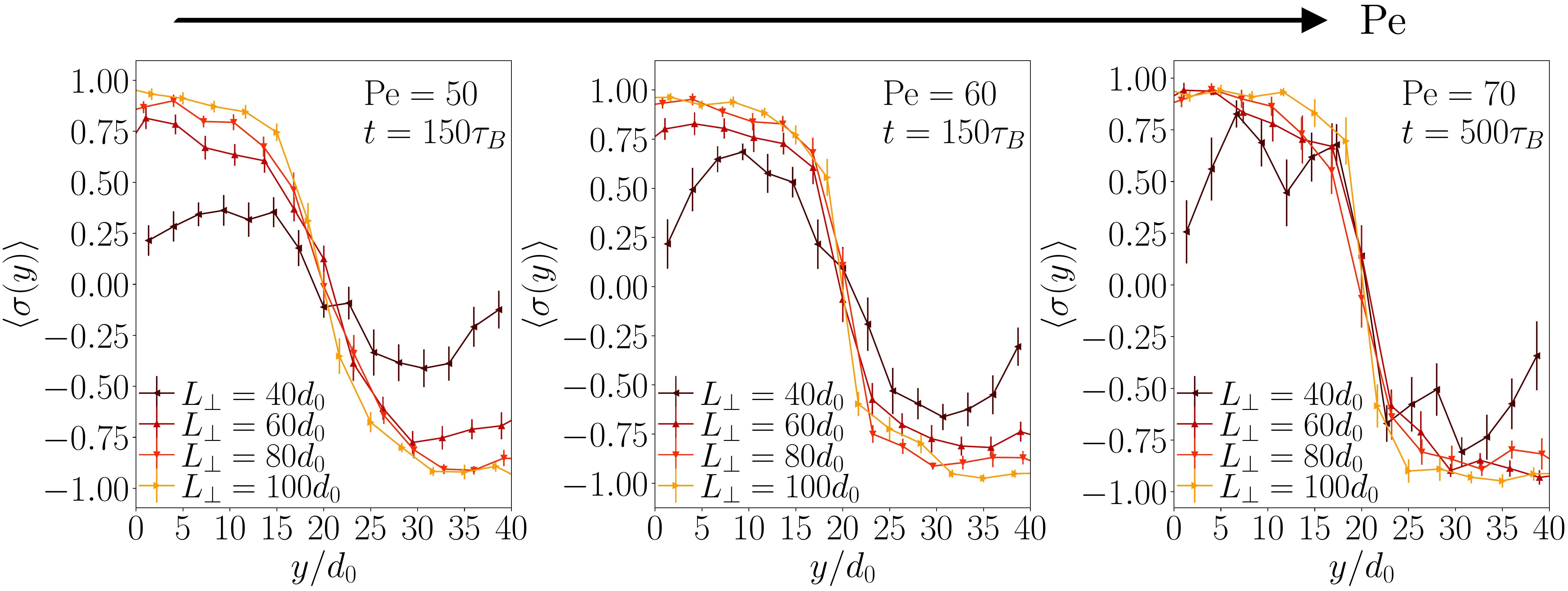}
    \caption{The charge density profiles with different aspect ratios, varying $\text{Pe}$ with $L_{\parallel} = 100d_0$.
    }
    \label{fig:S-Interface-Pe}
\end{figure*}

\subsection{Fitting for diffusive interface profile}
For small Pe, we expect the mixing for domains of red and blue particles 
 to be described by the diffusion equation: 
\begin{equation}
    \partial_{t} \langle\sigma(y, t)\rangle = D_{y} \partial_{y}^2 \langle\sigma(y, t)\rangle,
    \label{equ:diffy}
\end{equation}
where $D_{y}$ is a lateral diffusion constant.
The initial condition is (for $0\leq  y<L_y$) 
\begin{equation}
    \langle\sigma(y,  0)\rangle= \begin{cases}
     1 &  0\leq y< L_\perp/2 \\
     -1 & L_\perp/2 \leq  y < L_\perp \\
\end{cases}, 
\end{equation}
Since we consider a system with periodic boundary conditions, we extend the function  $\langle\sigma(y, t)\rangle$ to include all values of $y$ and we seek periodic solutions, such that
$\langle\sigma(y+ L_{\perp}, t)\rangle = \langle\sigma(y, t)\rangle$ for all $y, t$.  The corresponding solution to \eqref{equ:diffy} can be represented as
\begin{equation}
    \langle \sigma(y, t)\rangle = 
\sum_{k=-\infty}^{\infty} \left[ \erf\left( \frac{y-k L_\perp}{\sqrt{4D_{y} t}} \right)-\erf\left( \frac{y-\frac{2k-1}{2} L_{\perp}}{\sqrt{4D_{y}t}} \right) \right]
\end{equation}
For $0< y< L_\perp$ and moderate values of $t$, the contributions from large values of $k$ are negligible, so it is practical to truncate the sums.  In fact, for the relatively short times considered here, it is sufficient to approximate 
\begin{multline}
    \langle \sigma(y, t)\rangle = \erf\left( \frac{y}{\sqrt{4D_{y} t}} \right)-\erf\left( \frac{y-\tfrac12 L_{\perp}}{\sqrt{4D_{y} t}} \right)
      \\ + \erf\left( \frac{y - L_\perp}{\sqrt{4D_{y}t}} \right). 
\end{multline}
since all other contributions are small for $0\leq y < L_\perp$.
This form gives the fits shown in Fig.~\ref{fig:interface}.

\subsection{Interface mixing at different box shapes}

We explain in the main text that TJCs form at the interface between demixed red and blue domains, which causes the demixed state to break down in large systems.  As additional supporting evidence for this mechanism, we simulated systems with different aspect ratios: $L_{\parallel} = 100d_0$ and $L_{\perp} = 40d_0, 60d_0, 80d_0, 100d_0$ at $\text{Pe} = 50, 60, 70$.  This is the regime where diffusive mixing does not apply, and TJCs play an important role.

In Fig.~\ref{fig:S-Interface-time} we show the time-evolution of the interface profile at $\text{Pe} = 60$ and the mechanism described is consistent at different time steps.  Each interface profile is averaged over at least 20 simulations and the error bar for the density profile is taken to be the standard error of the average.  

The mixing effect depends very weakly on $L_\perp$, except for the smallest value ($L_\perp=40d_0$) which is small enough that a single TJC can affect the whole domain, and is no longer localised near an interface.  For the larger systems, this indicates that the mixing can be interpreted as an instability of a single interface between red and blue, initiated by a TJC.

Finally, Fig.~\ref{fig:S-Interface-Pe} shows snapshots of the interface profile at $\text{Pe} = 50, 60, 70$: the non-diffusive mixing mechanism is consistently observed across this range.

\end{appendix}

\bibliography{apssamp}% Produces the bibliography via BibTeX.

\providecommand{\noopsort}[1]{}\providecommand{\singleletter}[1]{#1}%
\begin{thebibliography}{72}%
\makeatletter
\providecommand \@ifxundefined [1]{%
 \@ifx{#1\undefined}
}%
\providecommand \@ifnum [1]{%
 \ifnum #1\expandafter \@firstoftwo
 \else \expandafter \@secondoftwo
 \fi
}%
\providecommand \@ifx [1]{%
 \ifx #1\expandafter \@firstoftwo
 \else \expandafter \@secondoftwo
 \fi
}%
\providecommand \natexlab [1]{#1}%
\providecommand \enquote  [1]{``#1''}%
\providecommand \bibnamefont  [1]{#1}%
\providecommand \bibfnamefont [1]{#1}%
\providecommand \citenamefont [1]{#1}%
\providecommand \href@noop [0]{\@secondoftwo}%
\providecommand \href [0]{\begingroup \@sanitize@url \@href}%
\providecommand \@href[1]{\@@startlink{#1}\@@href}%
\providecommand \@@href[1]{\endgroup#1\@@endlink}%
\providecommand \@sanitize@url [0]{\catcode `\\12\catcode `\$12\catcode `\&12\catcode `\#12\catcode `\^12\catcode `\_12\catcode `\%12\relax}%
\providecommand \@@startlink[1]{}%
\providecommand \@@endlink[0]{}%
\providecommand \url  [0]{\begingroup\@sanitize@url \@url }%
\providecommand \@url [1]{\endgroup\@href {#1}{\urlprefix }}%
\providecommand \urlprefix  [0]{URL }%
\providecommand \Eprint [0]{\href }%
\providecommand \doibase [0]{https://doi.org/}%
\providecommand \selectlanguage [0]{\@gobble}%
\providecommand \bibinfo  [0]{\@secondoftwo}%
\providecommand \bibfield  [0]{\@secondoftwo}%
\providecommand \translation [1]{[#1]}%
\providecommand \BibitemOpen [0]{}%
\providecommand \bibitemStop [0]{}%
\providecommand \bibitemNoStop [0]{.\EOS\space}%
\providecommand \EOS [0]{\spacefactor3000\relax}%
\providecommand \BibitemShut  [1]{\csname bibitem#1\endcsname}%
\let\auto@bib@innerbib\@empty
%</preamble>
\bibitem [{\citenamefont {Couzin}\ and\ \citenamefont {Krause}(2003)}]{couzin2003self}%
  \BibitemOpen
  \bibfield  {author} {\bibinfo {author} {\bibfnamefont {I.~D.}\ \bibnamefont {Couzin}}\ and\ \bibinfo {author} {\bibfnamefont {J.}~\bibnamefont {Krause}},\ }\href@noop {} {\bibfield  {journal} {\bibinfo  {journal} {Adv. Study Behav.}\ }\textbf {\bibinfo {volume} {32}},\ \bibinfo {pages} {10} (\bibinfo {year} {2003})}\BibitemShut {NoStop}%
\bibitem [{\citenamefont {Ballerini}\ \emph {et~al.}(2008)\citenamefont {Ballerini}, \citenamefont {Cabibbo}, \citenamefont {Candelier}, \citenamefont {Cavagna}, \citenamefont {Cisbani}, \citenamefont {Giardina}, \citenamefont {Orlandi}, \citenamefont {Parisi}, \citenamefont {Procaccini}, \citenamefont {Viale} \emph {et~al.}}]{ballerini2008empirical}%
  \BibitemOpen
  \bibfield  {author} {\bibinfo {author} {\bibfnamefont {M.}~\bibnamefont {Ballerini}}, \bibinfo {author} {\bibfnamefont {N.}~\bibnamefont {Cabibbo}}, \bibinfo {author} {\bibfnamefont {R.}~\bibnamefont {Candelier}}, \bibinfo {author} {\bibfnamefont {A.}~\bibnamefont {Cavagna}}, \bibinfo {author} {\bibfnamefont {E.}~\bibnamefont {Cisbani}}, \bibinfo {author} {\bibfnamefont {I.}~\bibnamefont {Giardina}}, \bibinfo {author} {\bibfnamefont {A.}~\bibnamefont {Orlandi}}, \bibinfo {author} {\bibfnamefont {G.}~\bibnamefont {Parisi}}, \bibinfo {author} {\bibfnamefont {A.}~\bibnamefont {Procaccini}}, \bibinfo {author} {\bibfnamefont {M.}~\bibnamefont {Viale}}, \emph {et~al.},\ }\href@noop {} {\bibfield  {journal} {\bibinfo  {journal} {Animal Behaviour}\ }\textbf {\bibinfo {volume} {76}},\ \bibinfo {pages} {201} (\bibinfo {year} {2008})}\BibitemShut {NoStop}%
\bibitem [{\citenamefont {Bialek}\ \emph {et~al.}(2012)\citenamefont {Bialek}, \citenamefont {Cavagna}, \citenamefont {Giardina}, \citenamefont {Mora}, \citenamefont {Silvestri}, \citenamefont {Viale},\ and\ \citenamefont {Walczak}}]{bialek2012statistical}%
  \BibitemOpen
  \bibfield  {author} {\bibinfo {author} {\bibfnamefont {W.}~\bibnamefont {Bialek}}, \bibinfo {author} {\bibfnamefont {A.}~\bibnamefont {Cavagna}}, \bibinfo {author} {\bibfnamefont {I.}~\bibnamefont {Giardina}}, \bibinfo {author} {\bibfnamefont {T.}~\bibnamefont {Mora}}, \bibinfo {author} {\bibfnamefont {E.}~\bibnamefont {Silvestri}}, \bibinfo {author} {\bibfnamefont {M.}~\bibnamefont {Viale}},\ and\ \bibinfo {author} {\bibfnamefont {A.~M.}\ \bibnamefont {Walczak}},\ }\href@noop {} {\bibfield  {journal} {\bibinfo  {journal} {Proc. Natl. Acad. Sci. USA}\ }\textbf {\bibinfo {volume} {109}},\ \bibinfo {pages} {4786} (\bibinfo {year} {2012})}\BibitemShut {NoStop}%
\bibitem [{\citenamefont {Sanchez}\ \emph {et~al.}(2012)\citenamefont {Sanchez}, \citenamefont {Chen}, \citenamefont {DeCamp}, \citenamefont {Heymann},\ and\ \citenamefont {Dogic}}]{sanchez2012spontaneous}%
  \BibitemOpen
  \bibfield  {author} {\bibinfo {author} {\bibfnamefont {T.}~\bibnamefont {Sanchez}}, \bibinfo {author} {\bibfnamefont {D.~T.}\ \bibnamefont {Chen}}, \bibinfo {author} {\bibfnamefont {S.~J.}\ \bibnamefont {DeCamp}}, \bibinfo {author} {\bibfnamefont {M.}~\bibnamefont {Heymann}},\ and\ \bibinfo {author} {\bibfnamefont {Z.}~\bibnamefont {Dogic}},\ }\href@noop {} {\bibfield  {journal} {\bibinfo  {journal} {Nature}\ }\textbf {\bibinfo {volume} {491}},\ \bibinfo {pages} {431} (\bibinfo {year} {2012})}\BibitemShut {NoStop}%
\bibitem [{\citenamefont {Doostmohammadi}\ \emph {et~al.}(2018)\citenamefont {Doostmohammadi}, \citenamefont {Ign{\'e}s-Mullol}, \citenamefont {Yeomans},\ and\ \citenamefont {Sagu{\'e}s}}]{doostmohammadi2018active}%
  \BibitemOpen
  \bibfield  {author} {\bibinfo {author} {\bibfnamefont {A.}~\bibnamefont {Doostmohammadi}}, \bibinfo {author} {\bibfnamefont {J.}~\bibnamefont {Ign{\'e}s-Mullol}}, \bibinfo {author} {\bibfnamefont {J.~M.}\ \bibnamefont {Yeomans}},\ and\ \bibinfo {author} {\bibfnamefont {F.}~\bibnamefont {Sagu{\'e}s}},\ }\href@noop {} {\bibfield  {journal} {\bibinfo  {journal} {Nat. Commun.}\ }\textbf {\bibinfo {volume} {9}},\ \bibinfo {pages} {3246} (\bibinfo {year} {2018})}\BibitemShut {NoStop}%
\bibitem [{\citenamefont {Tailleur}\ and\ \citenamefont {Cates}(2008)}]{tailleur2008statistical}%
  \BibitemOpen
  \bibfield  {author} {\bibinfo {author} {\bibfnamefont {J.}~\bibnamefont {Tailleur}}\ and\ \bibinfo {author} {\bibfnamefont {M.}~\bibnamefont {Cates}},\ }\href@noop {} {\bibfield  {journal} {\bibinfo  {journal} {Phys. Rev. Lett.}\ }\textbf {\bibinfo {volume} {100}},\ \bibinfo {pages} {218103} (\bibinfo {year} {2008})}\BibitemShut {NoStop}%
\bibitem [{\citenamefont {Thompson}\ \emph {et~al.}(2011)\citenamefont {Thompson}, \citenamefont {Tailleur}, \citenamefont {Cates},\ and\ \citenamefont {Blythe}}]{thompson2011lattice}%
  \BibitemOpen
  \bibfield  {author} {\bibinfo {author} {\bibfnamefont {A.~G.}\ \bibnamefont {Thompson}}, \bibinfo {author} {\bibfnamefont {J.}~\bibnamefont {Tailleur}}, \bibinfo {author} {\bibfnamefont {M.~E.}\ \bibnamefont {Cates}},\ and\ \bibinfo {author} {\bibfnamefont {R.~A.}\ \bibnamefont {Blythe}},\ }\href@noop {} {\bibfield  {journal} {\bibinfo  {journal} {J. Stat. Mech.: Theory Exp.}\ }\textbf {\bibinfo {volume} {2011}}\bibinfo  {number} { (02)},\ \bibinfo {pages} {P02029}}\BibitemShut {NoStop}%
\bibitem [{\citenamefont {Cates}\ and\ \citenamefont {Tailleur}(2015)}]{cates2015motility}%
  \BibitemOpen
\bibfield  {number} {  }\bibfield  {author} {\bibinfo {author} {\bibfnamefont {M.~E.}\ \bibnamefont {Cates}}\ and\ \bibinfo {author} {\bibfnamefont {J.}~\bibnamefont {Tailleur}},\ }\href@noop {} {\bibfield  {journal} {\bibinfo  {journal} {Annu. Rev. Condens. Matter Phys.}\ }\textbf {\bibinfo {volume} {6}},\ \bibinfo {pages} {219} (\bibinfo {year} {2015})}\BibitemShut {NoStop}%
\bibitem [{\citenamefont {Turing}(1952)}]{turing1952chemical}%
  \BibitemOpen
  \bibfield  {author} {\bibinfo {author} {\bibfnamefont {A.~M.}\ \bibnamefont {Turing}},\ }\href@noop {} {\bibfield  {journal} {\bibinfo  {journal} {Proc. R. Soc. London B}\ }\textbf {\bibinfo {volume} {237}},\ \bibinfo {pages} {37} (\bibinfo {year} {1952})}\BibitemShut {NoStop}%
\bibitem [{\citenamefont {Gray}\ and\ \citenamefont {Scott}(1983)}]{gray1983autocatalytic}%
  \BibitemOpen
  \bibfield  {author} {\bibinfo {author} {\bibfnamefont {P.}~\bibnamefont {Gray}}\ and\ \bibinfo {author} {\bibfnamefont {S.}~\bibnamefont {Scott}},\ }\href@noop {} {\bibfield  {journal} {\bibinfo  {journal} {Chem. Eng. Sci.}\ }\textbf {\bibinfo {volume} {38}},\ \bibinfo {pages} {29} (\bibinfo {year} {1983})}\BibitemShut {NoStop}%
\bibitem [{\citenamefont {Burger}\ \emph {et~al.}(2016)\citenamefont {Burger}, \citenamefont {Hittmeir}, \citenamefont {Ranetbauer},\ and\ \citenamefont {Wolfram}}]{burger2016lane}%
  \BibitemOpen
  \bibfield  {author} {\bibinfo {author} {\bibfnamefont {M.}~\bibnamefont {Burger}}, \bibinfo {author} {\bibfnamefont {S.}~\bibnamefont {Hittmeir}}, \bibinfo {author} {\bibfnamefont {H.}~\bibnamefont {Ranetbauer}},\ and\ \bibinfo {author} {\bibfnamefont {M.-T.}\ \bibnamefont {Wolfram}},\ }\href@noop {} {\bibfield  {journal} {\bibinfo  {journal} {SIAM J. Math. Anal.}\ }\textbf {\bibinfo {volume} {48}},\ \bibinfo {pages} {981} (\bibinfo {year} {2016})}\BibitemShut {NoStop}%
\bibitem [{\citenamefont {Dzubiella}\ \emph {et~al.}(2002)\citenamefont {Dzubiella}, \citenamefont {Hoffmann},\ and\ \citenamefont {L{\"o}wen}}]{dzubiella2002lane}%
  \BibitemOpen
  \bibfield  {author} {\bibinfo {author} {\bibfnamefont {J.}~\bibnamefont {Dzubiella}}, \bibinfo {author} {\bibfnamefont {G.}~\bibnamefont {Hoffmann}},\ and\ \bibinfo {author} {\bibfnamefont {H.}~\bibnamefont {L{\"o}wen}},\ }\href@noop {} {\bibfield  {journal} {\bibinfo  {journal} {Phys. Rev. E}\ }\textbf {\bibinfo {volume} {65}},\ \bibinfo {pages} {021402} (\bibinfo {year} {2002})}\BibitemShut {NoStop}%
\bibitem [{\citenamefont {Leunissen}\ \emph {et~al.}(2005)\citenamefont {Leunissen}, \citenamefont {Christova}, \citenamefont {Hynninen}, \citenamefont {Royall}, \citenamefont {Campbell}, \citenamefont {Imhof}, \citenamefont {Dijkstra}, \citenamefont {Van~Roij},\ and\ \citenamefont {Van~Blaaderen}}]{leunissen2005ionic}%
  \BibitemOpen
  \bibfield  {author} {\bibinfo {author} {\bibfnamefont {M.~E.}\ \bibnamefont {Leunissen}}, \bibinfo {author} {\bibfnamefont {C.~G.}\ \bibnamefont {Christova}}, \bibinfo {author} {\bibfnamefont {A.-P.}\ \bibnamefont {Hynninen}}, \bibinfo {author} {\bibfnamefont {C.~P.}\ \bibnamefont {Royall}}, \bibinfo {author} {\bibfnamefont {A.~I.}\ \bibnamefont {Campbell}}, \bibinfo {author} {\bibfnamefont {A.}~\bibnamefont {Imhof}}, \bibinfo {author} {\bibfnamefont {M.}~\bibnamefont {Dijkstra}}, \bibinfo {author} {\bibfnamefont {R.}~\bibnamefont {Van~Roij}},\ and\ \bibinfo {author} {\bibfnamefont {A.}~\bibnamefont {Van~Blaaderen}},\ }\href@noop {} {\bibfield  {journal} {\bibinfo  {journal} {Nature}\ }\textbf {\bibinfo {volume} {437}},\ \bibinfo {pages} {235} (\bibinfo {year} {2005})}\BibitemShut {NoStop}%
\bibitem [{\citenamefont {Vissers}\ \emph {et~al.}(2011{\natexlab{a}})\citenamefont {Vissers}, \citenamefont {Wysocki}, \citenamefont {Rex}, \citenamefont {L{\"o}wen}, \citenamefont {Royall}, \citenamefont {Imhof},\ and\ \citenamefont {van Blaaderen}}]{vissers2011lane}%
  \BibitemOpen
  \bibfield  {author} {\bibinfo {author} {\bibfnamefont {T.}~\bibnamefont {Vissers}}, \bibinfo {author} {\bibfnamefont {A.}~\bibnamefont {Wysocki}}, \bibinfo {author} {\bibfnamefont {M.}~\bibnamefont {Rex}}, \bibinfo {author} {\bibfnamefont {H.}~\bibnamefont {L{\"o}wen}}, \bibinfo {author} {\bibfnamefont {C.~P.}\ \bibnamefont {Royall}}, \bibinfo {author} {\bibfnamefont {A.}~\bibnamefont {Imhof}},\ and\ \bibinfo {author} {\bibfnamefont {A.}~\bibnamefont {van Blaaderen}},\ }\href@noop {} {\bibfield  {journal} {\bibinfo  {journal} {Soft Matter}\ }\textbf {\bibinfo {volume} {7}},\ \bibinfo {pages} {2352} (\bibinfo {year} {2011}{\natexlab{a}})}\BibitemShut {NoStop}%
\bibitem [{\citenamefont {S{\"u}tterlin}\ \emph {et~al.}(2009)\citenamefont {S{\"u}tterlin}, \citenamefont {Wysocki}, \citenamefont {Ivlev}, \citenamefont {R{\"a}th}, \citenamefont {Thomas}, \citenamefont {Rubin-Zuzic}, \citenamefont {Goedheer}, \citenamefont {Fortov}, \citenamefont {Lipaev}, \citenamefont {Molotkov} \emph {et~al.}}]{sutterlin2009dynamics}%
  \BibitemOpen
  \bibfield  {author} {\bibinfo {author} {\bibfnamefont {K.}~\bibnamefont {S{\"u}tterlin}}, \bibinfo {author} {\bibfnamefont {A.}~\bibnamefont {Wysocki}}, \bibinfo {author} {\bibfnamefont {A.}~\bibnamefont {Ivlev}}, \bibinfo {author} {\bibfnamefont {C.}~\bibnamefont {R{\"a}th}}, \bibinfo {author} {\bibfnamefont {H.}~\bibnamefont {Thomas}}, \bibinfo {author} {\bibfnamefont {M.}~\bibnamefont {Rubin-Zuzic}}, \bibinfo {author} {\bibfnamefont {W.}~\bibnamefont {Goedheer}}, \bibinfo {author} {\bibfnamefont {V.}~\bibnamefont {Fortov}}, \bibinfo {author} {\bibfnamefont {A.}~\bibnamefont {Lipaev}}, \bibinfo {author} {\bibfnamefont {V.}~\bibnamefont {Molotkov}}, \emph {et~al.},\ }\href@noop {} {\bibfield  {journal} {\bibinfo  {journal} {Phys. Rev. Lett.}\ }\textbf {\bibinfo {volume} {102}},\ \bibinfo {pages} {085003} (\bibinfo {year} {2009})}\BibitemShut {NoStop}%
\bibitem [{\citenamefont {Sutterlin}\ \emph {et~al.}(2009)\citenamefont {Sutterlin}, \citenamefont {Thomas}, \citenamefont {Ivlev}, \citenamefont {Morfill}, \citenamefont {Fortov}, \citenamefont {Lipaev}, \citenamefont {Molotkov}, \citenamefont {Petrov}, \citenamefont {Wysocki},\ and\ \citenamefont {Lowen}}]{sutterlin2009lane}%
  \BibitemOpen
  \bibfield  {author} {\bibinfo {author} {\bibfnamefont {K.~R.}\ \bibnamefont {Sutterlin}}, \bibinfo {author} {\bibfnamefont {H.~M.}\ \bibnamefont {Thomas}}, \bibinfo {author} {\bibfnamefont {A.~V.}\ \bibnamefont {Ivlev}}, \bibinfo {author} {\bibfnamefont {G.~E.}\ \bibnamefont {Morfill}}, \bibinfo {author} {\bibfnamefont {V.~E.}\ \bibnamefont {Fortov}}, \bibinfo {author} {\bibfnamefont {A.~M.}\ \bibnamefont {Lipaev}}, \bibinfo {author} {\bibfnamefont {V.~I.}\ \bibnamefont {Molotkov}}, \bibinfo {author} {\bibfnamefont {O.~F.}\ \bibnamefont {Petrov}}, \bibinfo {author} {\bibfnamefont {A.}~\bibnamefont {Wysocki}},\ and\ \bibinfo {author} {\bibfnamefont {H.}~\bibnamefont {Lowen}},\ }\href@noop {} {\bibfield  {journal} {\bibinfo  {journal} {IEEE Trans. Plasma Sci.}\ }\textbf {\bibinfo {volume} {38}},\ \bibinfo {pages} {861} (\bibinfo {year} {2009})}\BibitemShut {NoStop}%
\bibitem [{\citenamefont {Du}\ \emph {et~al.}(2012)\citenamefont {Du}, \citenamefont {S{\"u}tterlin}, \citenamefont {Jiang}, \citenamefont {R{\"a}th}, \citenamefont {Ivlev}, \citenamefont {Khrapak}, \citenamefont {Schwabe}, \citenamefont {Thomas}, \citenamefont {Fortov}, \citenamefont {Lipaev} \emph {et~al.}}]{du2012experimental}%
  \BibitemOpen
  \bibfield  {author} {\bibinfo {author} {\bibfnamefont {C.}~\bibnamefont {Du}}, \bibinfo {author} {\bibfnamefont {K.}~\bibnamefont {S{\"u}tterlin}}, \bibinfo {author} {\bibfnamefont {K.}~\bibnamefont {Jiang}}, \bibinfo {author} {\bibfnamefont {C.}~\bibnamefont {R{\"a}th}}, \bibinfo {author} {\bibfnamefont {A.}~\bibnamefont {Ivlev}}, \bibinfo {author} {\bibfnamefont {S.}~\bibnamefont {Khrapak}}, \bibinfo {author} {\bibfnamefont {M.}~\bibnamefont {Schwabe}}, \bibinfo {author} {\bibfnamefont {H.}~\bibnamefont {Thomas}}, \bibinfo {author} {\bibfnamefont {V.}~\bibnamefont {Fortov}}, \bibinfo {author} {\bibfnamefont {A.}~\bibnamefont {Lipaev}}, \emph {et~al.},\ }\href@noop {} {\bibfield  {journal} {\bibinfo  {journal} {New J. Phys.}\ }\textbf {\bibinfo {volume} {14}},\ \bibinfo {pages} {073058} (\bibinfo {year} {2012})}\BibitemShut {NoStop}%
\bibitem [{\citenamefont {Sarma}\ \emph {et~al.}(2020)\citenamefont {Sarma}, \citenamefont {Baruah},\ and\ \citenamefont {Ganesh}}]{sarma2020lane}%
  \BibitemOpen
  \bibfield  {author} {\bibinfo {author} {\bibfnamefont {U.}~\bibnamefont {Sarma}}, \bibinfo {author} {\bibfnamefont {S.}~\bibnamefont {Baruah}},\ and\ \bibinfo {author} {\bibfnamefont {R.}~\bibnamefont {Ganesh}},\ }\href@noop {} {\bibfield  {journal} {\bibinfo  {journal} {Phys. Plasmas}\ }\textbf {\bibinfo {volume} {27}},\ \bibinfo {pages} {012106} (\bibinfo {year} {2020})}\BibitemShut {NoStop}%
\bibitem [{\citenamefont {Lee}\ \emph {et~al.}(2014)\citenamefont {Lee}, \citenamefont {Kondrat}, \citenamefont {Oshanin},\ and\ \citenamefont {Kornyshev}}]{kondrat2014charging}%
  \BibitemOpen
  \bibfield  {author} {\bibinfo {author} {\bibfnamefont {A.~A.}\ \bibnamefont {Lee}}, \bibinfo {author} {\bibfnamefont {S.}~\bibnamefont {Kondrat}}, \bibinfo {author} {\bibfnamefont {G.}~\bibnamefont {Oshanin}},\ and\ \bibinfo {author} {\bibfnamefont {A.~A.}\ \bibnamefont {Kornyshev}},\ }\href@noop {} {\bibfield  {journal} {\bibinfo  {journal} {Nanotechnology}\ }\textbf {\bibinfo {volume} {25}},\ \bibinfo {pages} {315401} (\bibinfo {year} {2014})}\BibitemShut {NoStop}%
\bibitem [{\citenamefont {Helbing}\ and\ \citenamefont {Molnar}(1995)}]{helbing1995social}%
  \BibitemOpen
  \bibfield  {author} {\bibinfo {author} {\bibfnamefont {D.}~\bibnamefont {Helbing}}\ and\ \bibinfo {author} {\bibfnamefont {P.}~\bibnamefont {Molnar}},\ }\href@noop {} {\bibfield  {journal} {\bibinfo  {journal} {Phys. Rev. E}\ }\textbf {\bibinfo {volume} {51}},\ \bibinfo {pages} {4282} (\bibinfo {year} {1995})}\BibitemShut {NoStop}%
\bibitem [{\citenamefont {Isobe}\ \emph {et~al.}(2004)\citenamefont {Isobe}, \citenamefont {Adachi},\ and\ \citenamefont {Nagatani}}]{isobe2004experiment}%
  \BibitemOpen
  \bibfield  {author} {\bibinfo {author} {\bibfnamefont {M.}~\bibnamefont {Isobe}}, \bibinfo {author} {\bibfnamefont {T.}~\bibnamefont {Adachi}},\ and\ \bibinfo {author} {\bibfnamefont {T.}~\bibnamefont {Nagatani}},\ }\href@noop {} {\bibfield  {journal} {\bibinfo  {journal} {Physica A}\ }\textbf {\bibinfo {volume} {336}},\ \bibinfo {pages} {638} (\bibinfo {year} {2004})}\BibitemShut {NoStop}%
\bibitem [{\citenamefont {Karamouzas}\ \emph {et~al.}(2014)\citenamefont {Karamouzas}, \citenamefont {Skinner},\ and\ \citenamefont {Guy}}]{karamouzas2014universal}%
  \BibitemOpen
  \bibfield  {author} {\bibinfo {author} {\bibfnamefont {I.}~\bibnamefont {Karamouzas}}, \bibinfo {author} {\bibfnamefont {B.}~\bibnamefont {Skinner}},\ and\ \bibinfo {author} {\bibfnamefont {S.~J.}\ \bibnamefont {Guy}},\ }\href@noop {} {\bibfield  {journal} {\bibinfo  {journal} {Phys. Rev. Lett.}\ }\textbf {\bibinfo {volume} {113}},\ \bibinfo {pages} {238701} (\bibinfo {year} {2014})}\BibitemShut {NoStop}%
\bibitem [{\citenamefont {Oliveira}\ \emph {et~al.}(2016)\citenamefont {Oliveira}, \citenamefont {Vieira}, \citenamefont {Helbing}, \citenamefont {Andrade~Jr},\ and\ \citenamefont {Herrmann}}]{oliveira2016keep}%
  \BibitemOpen
  \bibfield  {author} {\bibinfo {author} {\bibfnamefont {C.~L.}\ \bibnamefont {Oliveira}}, \bibinfo {author} {\bibfnamefont {A.~P.}\ \bibnamefont {Vieira}}, \bibinfo {author} {\bibfnamefont {D.}~\bibnamefont {Helbing}}, \bibinfo {author} {\bibfnamefont {J.~S.}\ \bibnamefont {Andrade~Jr}},\ and\ \bibinfo {author} {\bibfnamefont {H.~J.}\ \bibnamefont {Herrmann}},\ }\href@noop {} {\bibfield  {journal} {\bibinfo  {journal} {Phys. Rev. X}\ }\textbf {\bibinfo {volume} {6}},\ \bibinfo {pages} {011003} (\bibinfo {year} {2016})}\BibitemShut {NoStop}%
\bibitem [{\citenamefont {Bacik}\ \emph {et~al.}(2023)\citenamefont {Bacik}, \citenamefont {Bacik},\ and\ \citenamefont {Rogers}}]{bacik2023lane}%
  \BibitemOpen
  \bibfield  {author} {\bibinfo {author} {\bibfnamefont {K.~A.}\ \bibnamefont {Bacik}}, \bibinfo {author} {\bibfnamefont {B.~S.}\ \bibnamefont {Bacik}},\ and\ \bibinfo {author} {\bibfnamefont {T.}~\bibnamefont {Rogers}},\ }\href@noop {} {\bibfield  {journal} {\bibinfo  {journal} {Science}\ }\textbf {\bibinfo {volume} {379}},\ \bibinfo {pages} {923} (\bibinfo {year} {2023})}\BibitemShut {NoStop}%
\bibitem [{\citenamefont {Couzin}\ and\ \citenamefont {Franks}(2003)}]{couzin2003self1}%
  \BibitemOpen
  \bibfield  {author} {\bibinfo {author} {\bibfnamefont {I.~D.}\ \bibnamefont {Couzin}}\ and\ \bibinfo {author} {\bibfnamefont {N.~R.}\ \bibnamefont {Franks}},\ }\href@noop {} {\bibfield  {journal} {\bibinfo  {journal} {Proc. R. Soc. London B}\ }\textbf {\bibinfo {volume} {270}},\ \bibinfo {pages} {139} (\bibinfo {year} {2003})}\BibitemShut {NoStop}%
\bibitem [{\citenamefont {Wysocki}\ and\ \citenamefont {L{\"o}wen}(2009)}]{wysocki2009oscillatory}%
  \BibitemOpen
  \bibfield  {author} {\bibinfo {author} {\bibfnamefont {A.}~\bibnamefont {Wysocki}}\ and\ \bibinfo {author} {\bibfnamefont {H.}~\bibnamefont {L{\"o}wen}},\ }\href@noop {} {\bibfield  {journal} {\bibinfo  {journal} {Phys. Rev. E}\ }\textbf {\bibinfo {volume} {79}},\ \bibinfo {pages} {041408} (\bibinfo {year} {2009})}\BibitemShut {NoStop}%
\bibitem [{\citenamefont {Vissers}\ \emph {et~al.}(2011{\natexlab{b}})\citenamefont {Vissers}, \citenamefont {van Blaaderen},\ and\ \citenamefont {Imhof}}]{vissers2011band}%
  \BibitemOpen
  \bibfield  {author} {\bibinfo {author} {\bibfnamefont {T.}~\bibnamefont {Vissers}}, \bibinfo {author} {\bibfnamefont {A.}~\bibnamefont {van Blaaderen}},\ and\ \bibinfo {author} {\bibfnamefont {A.}~\bibnamefont {Imhof}},\ }\href@noop {} {\bibfield  {journal} {\bibinfo  {journal} {Phys. Rev. Lett.}\ }\textbf {\bibinfo {volume} {106}},\ \bibinfo {pages} {228303} (\bibinfo {year} {2011}{\natexlab{b}})}\BibitemShut {NoStop}%
\bibitem [{\citenamefont {Li}\ \emph {et~al.}(2021)\citenamefont {Li}, \citenamefont {Wang}, \citenamefont {Shi}, \citenamefont {Gao}, \citenamefont {Shi}, \citenamefont {Woodward},\ and\ \citenamefont {Forsman}}]{li2021phase}%
  \BibitemOpen
  \bibfield  {author} {\bibinfo {author} {\bibfnamefont {B.}~\bibnamefont {Li}}, \bibinfo {author} {\bibfnamefont {Y.-L.}\ \bibnamefont {Wang}}, \bibinfo {author} {\bibfnamefont {G.}~\bibnamefont {Shi}}, \bibinfo {author} {\bibfnamefont {Y.}~\bibnamefont {Gao}}, \bibinfo {author} {\bibfnamefont {X.}~\bibnamefont {Shi}}, \bibinfo {author} {\bibfnamefont {C.~E.}\ \bibnamefont {Woodward}},\ and\ \bibinfo {author} {\bibfnamefont {J.}~\bibnamefont {Forsman}},\ }\href@noop {} {\bibfield  {journal} {\bibinfo  {journal} {ACS Nano}\ }\textbf {\bibinfo {volume} {15}},\ \bibinfo {pages} {2363} (\bibinfo {year} {2021})}\BibitemShut {NoStop}%
\bibitem [{\citenamefont {Rex}\ and\ \citenamefont {L{\"o}wen}(2008)}]{rex2008influence}%
  \BibitemOpen
  \bibfield  {author} {\bibinfo {author} {\bibfnamefont {M.}~\bibnamefont {Rex}}\ and\ \bibinfo {author} {\bibfnamefont {H.}~\bibnamefont {L{\"o}wen}},\ }\href@noop {} {\bibfield  {journal} {\bibinfo  {journal} {The European Physical Journal E}\ }\textbf {\bibinfo {volume} {26}},\ \bibinfo {pages} {143} (\bibinfo {year} {2008})}\BibitemShut {NoStop}%
\bibitem [{\citenamefont {Glanz}\ and\ \citenamefont {L{\"o}wen}(2012)}]{glanz2012nature}%
  \BibitemOpen
  \bibfield  {author} {\bibinfo {author} {\bibfnamefont {T.}~\bibnamefont {Glanz}}\ and\ \bibinfo {author} {\bibfnamefont {H.}~\bibnamefont {L{\"o}wen}},\ }\href@noop {} {\bibfield  {journal} {\bibinfo  {journal} {J. Phys.: Condens. Matter}\ }\textbf {\bibinfo {volume} {24}},\ \bibinfo {pages} {464114} (\bibinfo {year} {2012})}\BibitemShut {NoStop}%
\bibitem [{\citenamefont {Kohl}\ \emph {et~al.}(2012)\citenamefont {Kohl}, \citenamefont {Ivlev}, \citenamefont {Brandt}, \citenamefont {Morfill},\ and\ \citenamefont {L{\"o}wen}}]{kohl2012microscopic}%
  \BibitemOpen
  \bibfield  {author} {\bibinfo {author} {\bibfnamefont {M.}~\bibnamefont {Kohl}}, \bibinfo {author} {\bibfnamefont {A.~V.}\ \bibnamefont {Ivlev}}, \bibinfo {author} {\bibfnamefont {P.}~\bibnamefont {Brandt}}, \bibinfo {author} {\bibfnamefont {G.~E.}\ \bibnamefont {Morfill}},\ and\ \bibinfo {author} {\bibfnamefont {H.}~\bibnamefont {L{\"o}wen}},\ }\href@noop {} {\bibfield  {journal} {\bibinfo  {journal} {J. Phys.: Condens. Matter}\ }\textbf {\bibinfo {volume} {24}},\ \bibinfo {pages} {464115} (\bibinfo {year} {2012})}\BibitemShut {NoStop}%
\bibitem [{\citenamefont {Klymko}\ \emph {et~al.}(2016)\citenamefont {Klymko}, \citenamefont {Geissler},\ and\ \citenamefont {Whitelam}}]{klymko2016microscopic}%
  \BibitemOpen
  \bibfield  {author} {\bibinfo {author} {\bibfnamefont {K.}~\bibnamefont {Klymko}}, \bibinfo {author} {\bibfnamefont {P.~L.}\ \bibnamefont {Geissler}},\ and\ \bibinfo {author} {\bibfnamefont {S.}~\bibnamefont {Whitelam}},\ }\href@noop {} {\bibfield  {journal} {\bibinfo  {journal} {Phys. Rev. E}\ }\textbf {\bibinfo {volume} {94}},\ \bibinfo {pages} {022608} (\bibinfo {year} {2016})}\BibitemShut {NoStop}%
\bibitem [{\citenamefont {Poncet}\ \emph {et~al.}(2017)\citenamefont {Poncet}, \citenamefont {B{\'e}nichou}, \citenamefont {D{\'e}mery},\ and\ \citenamefont {Oshanin}}]{poncet2017universal}%
  \BibitemOpen
  \bibfield  {author} {\bibinfo {author} {\bibfnamefont {A.}~\bibnamefont {Poncet}}, \bibinfo {author} {\bibfnamefont {O.}~\bibnamefont {B{\'e}nichou}}, \bibinfo {author} {\bibfnamefont {V.}~\bibnamefont {D{\'e}mery}},\ and\ \bibinfo {author} {\bibfnamefont {G.}~\bibnamefont {Oshanin}},\ }\href@noop {} {\bibfield  {journal} {\bibinfo  {journal} {Phys. Rev. Lett.}\ }\textbf {\bibinfo {volume} {118}},\ \bibinfo {pages} {118002} (\bibinfo {year} {2017})}\BibitemShut {NoStop}%
\bibitem [{\citenamefont {Chakrabarti}\ \emph {et~al.}(2003)\citenamefont {Chakrabarti}, \citenamefont {Dzubiella},\ and\ \citenamefont {L{\"o}wen}}]{chakrabarti2003dynamical}%
  \BibitemOpen
  \bibfield  {author} {\bibinfo {author} {\bibfnamefont {J.}~\bibnamefont {Chakrabarti}}, \bibinfo {author} {\bibfnamefont {J.}~\bibnamefont {Dzubiella}},\ and\ \bibinfo {author} {\bibfnamefont {H.}~\bibnamefont {L{\"o}wen}},\ }\href@noop {} {\bibfield  {journal} {\bibinfo  {journal} {Europhys. Lett.}\ }\textbf {\bibinfo {volume} {61}},\ \bibinfo {pages} {415} (\bibinfo {year} {2003})}\BibitemShut {NoStop}%
\bibitem [{\citenamefont {Chakrabarti}\ \emph {et~al.}(2004)\citenamefont {Chakrabarti}, \citenamefont {Dzubiella},\ and\ \citenamefont {L{\"o}wen}}]{chakrabarti2004reentrance}%
  \BibitemOpen
  \bibfield  {author} {\bibinfo {author} {\bibfnamefont {J.}~\bibnamefont {Chakrabarti}}, \bibinfo {author} {\bibfnamefont {J.}~\bibnamefont {Dzubiella}},\ and\ \bibinfo {author} {\bibfnamefont {H.}~\bibnamefont {L{\"o}wen}},\ }\href@noop {} {\bibfield  {journal} {\bibinfo  {journal} {Phys. Rev. E}\ }\textbf {\bibinfo {volume} {70}},\ \bibinfo {pages} {012401} (\bibinfo {year} {2004})}\BibitemShut {NoStop}%
\bibitem [{\citenamefont {Reichhardt}\ and\ \citenamefont {Reichhardt}(2006)}]{reichhardt2006cooperative}%
  \BibitemOpen
  \bibfield  {author} {\bibinfo {author} {\bibfnamefont {C.}~\bibnamefont {Reichhardt}}\ and\ \bibinfo {author} {\bibfnamefont {C.~J.~O.}\ \bibnamefont {Reichhardt}},\ }\href@noop {} {\bibfield  {journal} {\bibinfo  {journal} {Phys. Rev. E}\ }\textbf {\bibinfo {volume} {74}},\ \bibinfo {pages} {011403} (\bibinfo {year} {2006})}\BibitemShut {NoStop}%
\bibitem [{\citenamefont {Vasilyev}\ \emph {et~al.}(2017)\citenamefont {Vasilyev}, \citenamefont {B{\'e}nichou}, \citenamefont {Mej{\'\i}a-Monasterio}, \citenamefont {Weeks},\ and\ \citenamefont {Oshanin}}]{vasilyev2017cooperative}%
  \BibitemOpen
  \bibfield  {author} {\bibinfo {author} {\bibfnamefont {O.~A.}\ \bibnamefont {Vasilyev}}, \bibinfo {author} {\bibfnamefont {O.}~\bibnamefont {B{\'e}nichou}}, \bibinfo {author} {\bibfnamefont {C.}~\bibnamefont {Mej{\'\i}a-Monasterio}}, \bibinfo {author} {\bibfnamefont {E.~R.}\ \bibnamefont {Weeks}},\ and\ \bibinfo {author} {\bibfnamefont {G.}~\bibnamefont {Oshanin}},\ }\href@noop {} {\bibfield  {journal} {\bibinfo  {journal} {Soft Matter}\ }\textbf {\bibinfo {volume} {13}},\ \bibinfo {pages} {7617} (\bibinfo {year} {2017})}\BibitemShut {NoStop}%
\bibitem [{\citenamefont {Schimansky-Geier}\ \emph {et~al.}(2021)\citenamefont {Schimansky-Geier}, \citenamefont {Lindner}, \citenamefont {Milster},\ and\ \citenamefont {Neiman}}]{schimansky2021demixing}%
  \BibitemOpen
  \bibfield  {author} {\bibinfo {author} {\bibfnamefont {L.}~\bibnamefont {Schimansky-Geier}}, \bibinfo {author} {\bibfnamefont {B.}~\bibnamefont {Lindner}}, \bibinfo {author} {\bibfnamefont {S.}~\bibnamefont {Milster}},\ and\ \bibinfo {author} {\bibfnamefont {A.~B.}\ \bibnamefont {Neiman}},\ }\href@noop {} {\bibfield  {journal} {\bibinfo  {journal} {Phys. Rev. E}\ }\textbf {\bibinfo {volume} {103}},\ \bibinfo {pages} {022113} (\bibinfo {year} {2021})}\BibitemShut {NoStop}%
\bibitem [{\citenamefont {Yu}\ \emph {et~al.}(2022)\citenamefont {Yu}, \citenamefont {Thijssen},\ and\ \citenamefont {Jack}}]{yu2022lane}%
  \BibitemOpen
  \bibfield  {author} {\bibinfo {author} {\bibfnamefont {H.}~\bibnamefont {Yu}}, \bibinfo {author} {\bibfnamefont {K.}~\bibnamefont {Thijssen}},\ and\ \bibinfo {author} {\bibfnamefont {R.~L.}\ \bibnamefont {Jack}},\ }\href {https://doi.org/10.1103/PhysRevE.106.024129} {\bibfield  {journal} {\bibinfo  {journal} {Phys. Rev. E}\ }\textbf {\bibinfo {volume} {106}},\ \bibinfo {pages} {024129} (\bibinfo {year} {2022})}\BibitemShut {NoStop}%
\bibitem [{\citenamefont {VanSaders}\ and\ \citenamefont {Vitelli}(2023)}]{vansaders2023informational}%
  \BibitemOpen
  \bibfield  {author} {\bibinfo {author} {\bibfnamefont {B.}~\bibnamefont {VanSaders}}\ and\ \bibinfo {author} {\bibfnamefont {V.}~\bibnamefont {Vitelli}},\ }\href@noop {} {\bibfield  {journal} {\bibinfo  {journal} {arXiv preprint arXiv:2302.07402}\ } (\bibinfo {year} {2023})}\BibitemShut {NoStop}%
\bibitem [{\citenamefont {Geigenfeind}\ \emph {et~al.}(2020)\citenamefont {Geigenfeind}, \citenamefont {de~las Heras},\ and\ \citenamefont {Schmidt}}]{geigenfeind2020superadiabatic}%
  \BibitemOpen
  \bibfield  {author} {\bibinfo {author} {\bibfnamefont {T.}~\bibnamefont {Geigenfeind}}, \bibinfo {author} {\bibfnamefont {D.}~\bibnamefont {de~las Heras}},\ and\ \bibinfo {author} {\bibfnamefont {M.}~\bibnamefont {Schmidt}},\ }\href@noop {} {\bibfield  {journal} {\bibinfo  {journal} {Commun. Phys.}\ }\textbf {\bibinfo {volume} {3}},\ \bibinfo {pages} {1} (\bibinfo {year} {2020})}\BibitemShut {NoStop}%
\bibitem [{SM()}]{SM}%
  \BibitemOpen
  \href@noop {} {\bibinfo {title} {see supplemental material for the supplementary movies of the steady state behaviors at different peclet number regimes and the mixing between the counterpropagating red and blue domains triggered by tjcs.}}\BibitemShut {Stop}%
\bibitem [{\citenamefont {Weeks}\ \emph {et~al.}(1971)\citenamefont {Weeks}, \citenamefont {Chandler},\ and\ \citenamefont {Andersen}}]{weeks1971role}%
  \BibitemOpen
  \bibfield  {author} {\bibinfo {author} {\bibfnamefont {J.~D.}\ \bibnamefont {Weeks}}, \bibinfo {author} {\bibfnamefont {D.}~\bibnamefont {Chandler}},\ and\ \bibinfo {author} {\bibfnamefont {H.~C.}\ \bibnamefont {Andersen}},\ }\href@noop {} {\bibfield  {journal} {\bibinfo  {journal} {J. Chem. Phys.}\ }\textbf {\bibinfo {volume} {54}},\ \bibinfo {pages} {5237} (\bibinfo {year} {1971})}\BibitemShut {NoStop}%
\bibitem [{\citenamefont {Thompson}\ \emph {et~al.}(2022)\citenamefont {Thompson}, \citenamefont {Aktulga}, \citenamefont {Berger}, \citenamefont {Bolintineanu}, \citenamefont {Brown}, \citenamefont {Crozier}, \citenamefont {in~'t Veld}, \citenamefont {Kohlmeyer}, \citenamefont {Moore}, \citenamefont {Nguyen}, \citenamefont {Shan}, \citenamefont {Stevens}, \citenamefont {Tranchida}, \citenamefont {Trott},\ and\ \citenamefont {Plimpton}}]{LAMMPS}%
  \BibitemOpen
  \bibfield  {author} {\bibinfo {author} {\bibfnamefont {A.~P.}\ \bibnamefont {Thompson}}, \bibinfo {author} {\bibfnamefont {H.~M.}\ \bibnamefont {Aktulga}}, \bibinfo {author} {\bibfnamefont {R.}~\bibnamefont {Berger}}, \bibinfo {author} {\bibfnamefont {D.~S.}\ \bibnamefont {Bolintineanu}}, \bibinfo {author} {\bibfnamefont {W.~M.}\ \bibnamefont {Brown}}, \bibinfo {author} {\bibfnamefont {P.~S.}\ \bibnamefont {Crozier}}, \bibinfo {author} {\bibfnamefont {P.~J.}\ \bibnamefont {in~'t Veld}}, \bibinfo {author} {\bibfnamefont {A.}~\bibnamefont {Kohlmeyer}}, \bibinfo {author} {\bibfnamefont {S.~G.}\ \bibnamefont {Moore}}, \bibinfo {author} {\bibfnamefont {T.~D.}\ \bibnamefont {Nguyen}}, \bibinfo {author} {\bibfnamefont {R.}~\bibnamefont {Shan}}, \bibinfo {author} {\bibfnamefont {M.~J.}\ \bibnamefont {Stevens}}, \bibinfo {author} {\bibfnamefont {J.}~\bibnamefont {Tranchida}}, \bibinfo {author} {\bibfnamefont {C.}~\bibnamefont {Trott}},\ and\ \bibinfo {author} {\bibfnamefont {S.~J.}\ \bibnamefont {Plimpton}},\ }\href
  {https://doi.org/10.1016/j.cpc.2021.108171} {\bibfield  {journal} {\bibinfo  {journal} {Comp. Phys. Comm.}\ }\textbf {\bibinfo {volume} {271}},\ \bibinfo {pages} {108171} (\bibinfo {year} {2022})}\BibitemShut {NoStop}%
\bibitem [{\citenamefont {Korniss}\ \emph {et~al.}(1995)\citenamefont {Korniss}, \citenamefont {Schmittmann},\ and\ \citenamefont {Zia}}]{korniss1995novel}%
  \BibitemOpen
  \bibfield  {author} {\bibinfo {author} {\bibfnamefont {G.}~\bibnamefont {Korniss}}, \bibinfo {author} {\bibfnamefont {B.}~\bibnamefont {Schmittmann}},\ and\ \bibinfo {author} {\bibfnamefont {R.}~\bibnamefont {Zia}},\ }\href@noop {} {\bibfield  {journal} {\bibinfo  {journal} {Europhys. Lett.}\ }\textbf {\bibinfo {volume} {32}},\ \bibinfo {pages} {49} (\bibinfo {year} {1995})}\BibitemShut {NoStop}%
\bibitem [{\citenamefont {Korniss}\ \emph {et~al.}(1997)\citenamefont {Korniss}, \citenamefont {Schmittmann},\ and\ \citenamefont {Zia}}]{korniss1997nonequilibrium}%
  \BibitemOpen
  \bibfield  {author} {\bibinfo {author} {\bibfnamefont {G.}~\bibnamefont {Korniss}}, \bibinfo {author} {\bibfnamefont {B.}~\bibnamefont {Schmittmann}},\ and\ \bibinfo {author} {\bibfnamefont {R.}~\bibnamefont {Zia}},\ }\href@noop {} {\bibfield  {journal} {\bibinfo  {journal} {J. Stat. Phys.}\ }\textbf {\bibinfo {volume} {86}},\ \bibinfo {pages} {721} (\bibinfo {year} {1997})}\BibitemShut {NoStop}%
\bibitem [{\citenamefont {D{\'e}mery}\ and\ \citenamefont {Dean}(2016)}]{demery2016conductivity}%
  \BibitemOpen
  \bibfield  {author} {\bibinfo {author} {\bibfnamefont {V.}~\bibnamefont {D{\'e}mery}}\ and\ \bibinfo {author} {\bibfnamefont {D.~S.}\ \bibnamefont {Dean}},\ }\href@noop {} {\bibfield  {journal} {\bibinfo  {journal} {J. Stat. Mech.: Theory Exp.}\ }\textbf {\bibinfo {volume} {2016}}\bibinfo  {number} { (2)},\ \bibinfo {pages} {023106}}\BibitemShut {NoStop}%
\bibitem [{\citenamefont {Frusawa}(2022)}]{frusawa2022stochastic}%
  \BibitemOpen
\bibfield  {number} {  }\bibfield  {author} {\bibinfo {author} {\bibfnamefont {H.}~\bibnamefont {Frusawa}},\ }\href@noop {} {\bibfield  {journal} {\bibinfo  {journal} {Entropy}\ }\textbf {\bibinfo {volume} {24}},\ \bibinfo {pages} {500} (\bibinfo {year} {2022})}\BibitemShut {NoStop}%
\bibitem [{\citenamefont {Schmittmann}\ and\ \citenamefont {Zia}(1995)}]{SCHMITTMANN1995alt2}%
  \BibitemOpen
  \bibfield  {author} {\bibinfo {author} {\bibfnamefont {B.}~\bibnamefont {Schmittmann}}\ and\ \bibinfo {author} {\bibfnamefont {R.}~\bibnamefont {Zia}},\ }in\ \href@noop {} {\emph {\bibinfo {booktitle} {Statistical Mechanics of Driven Diffusive System}}},\ \bibinfo {series} {Phase Transitions and Critical Phenomena}, Vol.~\bibinfo {volume} {17},\ \bibinfo {editor} {edited by\ \bibinfo {editor} {\bibfnamefont {B.}~\bibnamefont {Schmittmann}}\ and\ \bibinfo {editor} {\bibfnamefont {R.}~\bibnamefont {Zia}}}\ (\bibinfo  {publisher} {Academic Press},\ \bibinfo {year} {1995})\ pp.\ \bibinfo {pages} {3--214}\BibitemShut {NoStop}%
\bibitem [{\citenamefont {Hansen}\ and\ \citenamefont {McDonald}(2013)}]{hansen2013theory}%
  \BibitemOpen
  \bibfield  {author} {\bibinfo {author} {\bibfnamefont {J.-P.}\ \bibnamefont {Hansen}}\ and\ \bibinfo {author} {\bibfnamefont {I.~R.}\ \bibnamefont {McDonald}},\ }\href@noop {} {\emph {\bibinfo {title} {Theory of simple liquids: with applications to soft matter}}}\ (\bibinfo  {publisher} {Academic press},\ \bibinfo {year} {2013})\BibitemShut {NoStop}%
\bibitem [{\citenamefont {Crooks}\ and\ \citenamefont {Chandler}(1997)}]{crooks1997gaussian}%
  \BibitemOpen
  \bibfield  {author} {\bibinfo {author} {\bibfnamefont {G.~E.}\ \bibnamefont {Crooks}}\ and\ \bibinfo {author} {\bibfnamefont {D.}~\bibnamefont {Chandler}},\ }\href@noop {} {\bibfield  {journal} {\bibinfo  {journal} {Phys. Rev. E}\ }\textbf {\bibinfo {volume} {56}},\ \bibinfo {pages} {4217} (\bibinfo {year} {1997})}\BibitemShut {NoStop}%
\bibitem [{\citenamefont {Del~Junco}\ \emph {et~al.}(2018)\citenamefont {Del~Junco}, \citenamefont {Tociu},\ and\ \citenamefont {Vaikuntanathan}}]{del2018energy}%
  \BibitemOpen
  \bibfield  {author} {\bibinfo {author} {\bibfnamefont {C.}~\bibnamefont {Del~Junco}}, \bibinfo {author} {\bibfnamefont {L.}~\bibnamefont {Tociu}},\ and\ \bibinfo {author} {\bibfnamefont {S.}~\bibnamefont {Vaikuntanathan}},\ }\href@noop {} {\bibfield  {journal} {\bibinfo  {journal} {Proc. Natl. Acad. Sci. USA}\ }\textbf {\bibinfo {volume} {115}},\ \bibinfo {pages} {3569} (\bibinfo {year} {2018})}\BibitemShut {NoStop}%
\bibitem [{\citenamefont {Omar}\ \emph {et~al.}(2021)\citenamefont {Omar}, \citenamefont {Klymko}, \citenamefont {GrandPre},\ and\ \citenamefont {Geissler}}]{omar2021phase}%
  \BibitemOpen
  \bibfield  {author} {\bibinfo {author} {\bibfnamefont {A.~K.}\ \bibnamefont {Omar}}, \bibinfo {author} {\bibfnamefont {K.}~\bibnamefont {Klymko}}, \bibinfo {author} {\bibfnamefont {T.}~\bibnamefont {GrandPre}},\ and\ \bibinfo {author} {\bibfnamefont {P.~L.}\ \bibnamefont {Geissler}},\ }\href@noop {} {\bibfield  {journal} {\bibinfo  {journal} {Phys. Rev. Lett.}\ }\textbf {\bibinfo {volume} {126}},\ \bibinfo {pages} {188002} (\bibinfo {year} {2021})}\BibitemShut {NoStop}%
\bibitem [{\citenamefont {Dutta}\ and\ \citenamefont {Chakrabarti}(2016)}]{dutta2016anomalous}%
  \BibitemOpen
  \bibfield  {author} {\bibinfo {author} {\bibfnamefont {S.}~\bibnamefont {Dutta}}\ and\ \bibinfo {author} {\bibfnamefont {J.}~\bibnamefont {Chakrabarti}},\ }\href@noop {} {\bibfield  {journal} {\bibinfo  {journal} {Europhys. Lett.}\ }\textbf {\bibinfo {volume} {116}},\ \bibinfo {pages} {38001} (\bibinfo {year} {2016})}\BibitemShut {NoStop}%
\bibitem [{\citenamefont {Dutta}\ and\ \citenamefont {Chakrabarti}(2018)}]{dutta2018transient}%
  \BibitemOpen
  \bibfield  {author} {\bibinfo {author} {\bibfnamefont {S.}~\bibnamefont {Dutta}}\ and\ \bibinfo {author} {\bibfnamefont {J.}~\bibnamefont {Chakrabarti}},\ }\href@noop {} {\bibfield  {journal} {\bibinfo  {journal} {Soft Matter}\ }\textbf {\bibinfo {volume} {14}},\ \bibinfo {pages} {4477} (\bibinfo {year} {2018})}\BibitemShut {NoStop}%
\bibitem [{\citenamefont {Dutta}\ and\ \citenamefont {Chakrabarti}(2020)}]{dutta2020length}%
  \BibitemOpen
  \bibfield  {author} {\bibinfo {author} {\bibfnamefont {S.}~\bibnamefont {Dutta}}\ and\ \bibinfo {author} {\bibfnamefont {J.}~\bibnamefont {Chakrabarti}},\ }\href@noop {} {\bibfield  {journal} {\bibinfo  {journal} {Phys. Chem. Chem. Phys.}\ }\textbf {\bibinfo {volume} {22}},\ \bibinfo {pages} {17731} (\bibinfo {year} {2020})}\BibitemShut {NoStop}%
\bibitem [{\citenamefont {Del~Junco}\ and\ \citenamefont {Vaikuntanathan}(2019)}]{del2019interface}%
  \BibitemOpen
  \bibfield  {author} {\bibinfo {author} {\bibfnamefont {C.}~\bibnamefont {Del~Junco}}\ and\ \bibinfo {author} {\bibfnamefont {S.}~\bibnamefont {Vaikuntanathan}},\ }\href@noop {} {\bibfield  {journal} {\bibinfo  {journal} {J. Chem. Phys.}\ }\textbf {\bibinfo {volume} {150}},\ \bibinfo {pages} {094708} (\bibinfo {year} {2019})}\BibitemShut {NoStop}%
\bibitem [{\citenamefont {Dean}\ \emph {et~al.}(2020)\citenamefont {Dean}, \citenamefont {Gersberg},\ and\ \citenamefont {Holdsworth}}]{dean2020effect}%
  \BibitemOpen
  \bibfield  {author} {\bibinfo {author} {\bibfnamefont {D.~S.}\ \bibnamefont {Dean}}, \bibinfo {author} {\bibfnamefont {P.}~\bibnamefont {Gersberg}},\ and\ \bibinfo {author} {\bibfnamefont {P.~C.}\ \bibnamefont {Holdsworth}},\ }\href@noop {} {\bibfield  {journal} {\bibinfo  {journal} {J. Stat. Mech.: Theory Exp.}\ }\textbf {\bibinfo {volume} {2020}}\bibinfo  {number} { (3)},\ \bibinfo {pages} {033206}}\BibitemShut {NoStop}%
\bibitem [{\citenamefont {Bouchet}\ \emph {et~al.}(2019)\citenamefont {Bouchet}, \citenamefont {Rolland},\ and\ \citenamefont {Simonnet}}]{bouchet2019rare}%
  \BibitemOpen
\bibfield  {number} {  }\bibfield  {author} {\bibinfo {author} {\bibfnamefont {F.}~\bibnamefont {Bouchet}}, \bibinfo {author} {\bibfnamefont {J.}~\bibnamefont {Rolland}},\ and\ \bibinfo {author} {\bibfnamefont {E.}~\bibnamefont {Simonnet}},\ }\href@noop {} {\bibfield  {journal} {\bibinfo  {journal} {Phys. Rev. Lett.}\ }\textbf {\bibinfo {volume} {122}},\ \bibinfo {pages} {074502} (\bibinfo {year} {2019})}\BibitemShut {NoStop}%
\bibitem [{\citenamefont {Helbing}\ \emph {et~al.}(2000)\citenamefont {Helbing}, \citenamefont {Farkas},\ and\ \citenamefont {Vicsek}}]{helbing2000freezing}%
  \BibitemOpen
  \bibfield  {author} {\bibinfo {author} {\bibfnamefont {D.}~\bibnamefont {Helbing}}, \bibinfo {author} {\bibfnamefont {I.~J.}\ \bibnamefont {Farkas}},\ and\ \bibinfo {author} {\bibfnamefont {T.}~\bibnamefont {Vicsek}},\ }\href@noop {} {\bibfield  {journal} {\bibinfo  {journal} {Phys. Rev. Lett.}\ }\textbf {\bibinfo {volume} {84}},\ \bibinfo {pages} {1240} (\bibinfo {year} {2000})}\BibitemShut {NoStop}%
\bibitem [{\citenamefont {Wensink}\ and\ \citenamefont {L{\"o}wen}(2012)}]{wensink2012emergent}%
  \BibitemOpen
  \bibfield  {author} {\bibinfo {author} {\bibfnamefont {H.}~\bibnamefont {Wensink}}\ and\ \bibinfo {author} {\bibfnamefont {H.}~\bibnamefont {L{\"o}wen}},\ }\href@noop {} {\bibfield  {journal} {\bibinfo  {journal} {J. Phys.: Condens. Matter}\ }\textbf {\bibinfo {volume} {24}},\ \bibinfo {pages} {464130} (\bibinfo {year} {2012})}\BibitemShut {NoStop}%
\bibitem [{\citenamefont {Farrell}\ \emph {et~al.}(2012)\citenamefont {Farrell}, \citenamefont {Marchetti}, \citenamefont {Marenduzzo},\ and\ \citenamefont {Tailleur}}]{Farrell2012}%
  \BibitemOpen
  \bibfield  {author} {\bibinfo {author} {\bibfnamefont {F.~D.~C.}\ \bibnamefont {Farrell}}, \bibinfo {author} {\bibfnamefont {M.~C.}\ \bibnamefont {Marchetti}}, \bibinfo {author} {\bibfnamefont {D.}~\bibnamefont {Marenduzzo}},\ and\ \bibinfo {author} {\bibfnamefont {J.}~\bibnamefont {Tailleur}},\ }\href {https://doi.org/10.1103/PhysRevLett.108.248101} {\bibfield  {journal} {\bibinfo  {journal} {Phys. Rev. Lett.}\ }\textbf {\bibinfo {volume} {108}},\ \bibinfo {pages} {248101} (\bibinfo {year} {2012})}\BibitemShut {NoStop}%
\bibitem [{\citenamefont {Bain}\ and\ \citenamefont {Bartolo}(2017)}]{bain2017critical}%
  \BibitemOpen
  \bibfield  {author} {\bibinfo {author} {\bibfnamefont {N.}~\bibnamefont {Bain}}\ and\ \bibinfo {author} {\bibfnamefont {D.}~\bibnamefont {Bartolo}},\ }\href@noop {} {\bibfield  {journal} {\bibinfo  {journal} {Nat. Commun.}\ }\textbf {\bibinfo {volume} {8}},\ \bibinfo {pages} {15969} (\bibinfo {year} {2017})}\BibitemShut {NoStop}%
\bibitem [{\citenamefont {B{\"a}r}\ \emph {et~al.}(2020)\citenamefont {B{\"a}r}, \citenamefont {Gro{\ss}mann}, \citenamefont {Heidenreich},\ and\ \citenamefont {Peruani}}]{Bar2020}%
  \BibitemOpen
  \bibfield  {author} {\bibinfo {author} {\bibfnamefont {M.}~\bibnamefont {B{\"a}r}}, \bibinfo {author} {\bibfnamefont {R.}~\bibnamefont {Gro{\ss}mann}}, \bibinfo {author} {\bibfnamefont {S.}~\bibnamefont {Heidenreich}},\ and\ \bibinfo {author} {\bibfnamefont {F.}~\bibnamefont {Peruani}},\ }\href@noop {} {\bibfield  {journal} {\bibinfo  {journal} {Annu. Rev. Condens. Matter Phys.}\ }\textbf {\bibinfo {volume} {11}},\ \bibinfo {pages} {441} (\bibinfo {year} {2020})}\BibitemShut {NoStop}%
\bibitem [{\citenamefont {Stopper}\ and\ \citenamefont {Roth}(2018)}]{Stopper2018}%
  \BibitemOpen
  \bibfield  {author} {\bibinfo {author} {\bibfnamefont {D.}~\bibnamefont {Stopper}}\ and\ \bibinfo {author} {\bibfnamefont {R.}~\bibnamefont {Roth}},\ }\href {https://doi.org/10.1103/PhysRevE.97.062602} {\bibfield  {journal} {\bibinfo  {journal} {Phys. Rev. E}\ }\textbf {\bibinfo {volume} {97}},\ \bibinfo {pages} {062602} (\bibinfo {year} {2018})}\BibitemShut {NoStop}%
\bibitem [{\citenamefont {Solon}\ \emph {et~al.}(2022)\citenamefont {Solon}, \citenamefont {Chat{\'e}}, \citenamefont {Toner},\ and\ \citenamefont {Tailleur}}]{solon2022susceptibility}%
  \BibitemOpen
  \bibfield  {author} {\bibinfo {author} {\bibfnamefont {A.}~\bibnamefont {Solon}}, \bibinfo {author} {\bibfnamefont {H.}~\bibnamefont {Chat{\'e}}}, \bibinfo {author} {\bibfnamefont {J.}~\bibnamefont {Toner}},\ and\ \bibinfo {author} {\bibfnamefont {J.}~\bibnamefont {Tailleur}},\ }\href@noop {} {\bibfield  {journal} {\bibinfo  {journal} {Phys. Rev. Lett.}\ }\textbf {\bibinfo {volume} {128}},\ \bibinfo {pages} {208004} (\bibinfo {year} {2022})}\BibitemShut {NoStop}%
\bibitem [{\citenamefont {Anderson}\ \emph {et~al.}(2023)\citenamefont {Anderson}, \citenamefont {Goldsztein},\ and\ \citenamefont {Fernandez-Nieves}}]{anderson2023ant}%
  \BibitemOpen
  \bibfield  {author} {\bibinfo {author} {\bibfnamefont {C.}~\bibnamefont {Anderson}}, \bibinfo {author} {\bibfnamefont {G.}~\bibnamefont {Goldsztein}},\ and\ \bibinfo {author} {\bibfnamefont {A.}~\bibnamefont {Fernandez-Nieves}},\ }\href@noop {} {\bibfield  {journal} {\bibinfo  {journal} {Sci. Adv.}\ }\textbf {\bibinfo {volume} {9}},\ \bibinfo {pages} {eadd0635} (\bibinfo {year} {2023})}\BibitemShut {NoStop}%
\bibitem [{\citenamefont {Zhang}\ and\ \citenamefont {Fodor}(2023)}]{zhang2022arxiv}%
  \BibitemOpen
  \bibfield  {author} {\bibinfo {author} {\bibfnamefont {Y.}~\bibnamefont {Zhang}}\ and\ \bibinfo {author} {\bibfnamefont {{\'E}.}~\bibnamefont {Fodor}},\ }\href@noop {} {\bibfield  {journal} {\bibinfo  {journal} {Phys. Rev. Lett.}\ }\textbf {\bibinfo {volume} {131}},\ \bibinfo {pages} {238302} (\bibinfo {year} {2023})}\BibitemShut {NoStop}%
\bibitem [{\citenamefont {Keta}\ \emph {et~al.}()\citenamefont {Keta}, \citenamefont {Klamser}, \citenamefont {Jack},\ and\ \citenamefont {Berthier}}]{keta2023arxiv}%
  \BibitemOpen
  \bibfield  {author} {\bibinfo {author} {\bibfnamefont {Y.-E.}\ \bibnamefont {Keta}}, \bibinfo {author} {\bibfnamefont {J.~U.}\ \bibnamefont {Klamser}}, \bibinfo {author} {\bibfnamefont {R.~L.}\ \bibnamefont {Jack}},\ and\ \bibinfo {author} {\bibfnamefont {L.}~\bibnamefont {Berthier}},\ }\href@noop {} {}\bibinfo {note} {ArXiv:2306.07172}\BibitemShut {NoStop}%
\bibitem [{\citenamefont {Chennakesavalu}\ and\ \citenamefont {Rotskoff}(2021)}]{chennakesavalu2021probing}%
  \BibitemOpen
  \bibfield  {author} {\bibinfo {author} {\bibfnamefont {S.}~\bibnamefont {Chennakesavalu}}\ and\ \bibinfo {author} {\bibfnamefont {G.~M.}\ \bibnamefont {Rotskoff}},\ }\href@noop {} {\bibfield  {journal} {\bibinfo  {journal} {J. Chem. Phys.}\ }\textbf {\bibinfo {volume} {155}},\ \bibinfo {pages} {194114} (\bibinfo {year} {2021})}\BibitemShut {NoStop}%
\bibitem [{\citenamefont {Berthier}\ \emph {et~al.}(2019)\citenamefont {Berthier}, \citenamefont {Flenner},\ and\ \citenamefont {Szamel}}]{berthier2019glassy}%
  \BibitemOpen
  \bibfield  {author} {\bibinfo {author} {\bibfnamefont {L.}~\bibnamefont {Berthier}}, \bibinfo {author} {\bibfnamefont {E.}~\bibnamefont {Flenner}},\ and\ \bibinfo {author} {\bibfnamefont {G.}~\bibnamefont {Szamel}},\ }\href@noop {} {\bibfield  {journal} {\bibinfo  {journal} {J. Chem. Phys.}\ }\textbf {\bibinfo {volume} {150}},\ \bibinfo {pages} {200901} (\bibinfo {year} {2019})}\BibitemShut {NoStop}%
\bibitem [{\citenamefont {Keta}\ \emph {et~al.}(2022)\citenamefont {Keta}, \citenamefont {Jack},\ and\ \citenamefont {Berthier}}]{keta2022disordered}%
  \BibitemOpen
  \bibfield  {author} {\bibinfo {author} {\bibfnamefont {Y.-E.}\ \bibnamefont {Keta}}, \bibinfo {author} {\bibfnamefont {R.~L.}\ \bibnamefont {Jack}},\ and\ \bibinfo {author} {\bibfnamefont {L.}~\bibnamefont {Berthier}},\ }\href@noop {} {\bibfield  {journal} {\bibinfo  {journal} {Phys. Rev. Lett.}\ }\textbf {\bibinfo {volume} {129}},\ \bibinfo {pages} {048002} (\bibinfo {year} {2022})}\BibitemShut {NoStop}%
\end{thebibliography}%

\end{document}